\documentclass{article}

\usepackage[utf8]{inputenc}
\usepackage{arxiv}
\usepackage{graphicx} 
\usepackage{amsmath}
\usepackage{amssymb}

\newcommand{\bss}{\boldsymbol{s}}

\newcommand{\bse}{\boldsymbol{e}}
\newcommand{\bsn}{\boldsymbol{n}}

\newcommand{\bsx}{\boldsymbol{x}}

\newcommand{\bsJ}{\boldsymbol{J}}

\newcommand{\bsu}{\boldsymbol{u}}

\newcommand{\bsomega}{\boldsymbol{\omega}}

\newcommand{\pp}{\partial}

\usepackage{xcolor}
\definecolor{dodgerblue}{rgb}{0.11,0.56,1}

\title{On the effect of airfoil geometry on extreme vortex-gust encounters}

\author{
    Barbara Lopez-Doriga\thanks{Email for correspondence: bldoriga@ucla.edu}, \  Anya R. M. Jones, \ and Kunihiko Taira \\ \\
    Department of Mechanical and Aerospace Engineering, University of California, Los Angeles, CA 90095, USA 
}

\begin{document}
\maketitle
\begin{abstract}
Historically, investigations on gust encounters have been limited to thin airfoils. In this work, we examine vortex-gust encounters by a family of airfoils at a chord-based Reynolds number $Re_c=100$, which includes variations in the gust ratio, initial gust position, gust radius, angle of attack, airfoil thickness, and airfoil camber. We examine differences in the flow fields, lift-element distributions, and aerodynamic responses across several airfoil-gust interactions. We observe a large deviation of the flow fields and aerodynamic responses with respect to the baseline flows for increasing gust ratios and gust sizes. The initial position of the vortex gust influences the magnitude of the velocity gradients observed near the leading edge, effectively heightening or mitigating the amplitude of the lift response. Moreover, the lift fluctuation increases with the angle of attack until it flattens around $10^\circ$, reminiscent of an unsteady stall-like regime. Furthermore, we report a decrease in the amplitude of the gust-induced lift fluctuations for thicker airfoils, which we attribute to a decrease in the vorticity production levels from the leading edge. The exploration of a sensitive subset of the parameter space uncovers relevant trends, shedding light on regions that have received limited attention in past studies, with special focus on the influence of airfoil geometry. 

\end{abstract}

\section{Introduction}
\label{sec:intro}
The development and improvement of small-scale aerial vehicles has become increasingly important, as they can perform a wide array of assignments safely and cost-effectively. A few of these applications include agriculture and medical supply deliveries to remote locations \cite{Panagiotou2020drones}, infrastructure inspections and urban mapping \cite{Shakhatreh2019}, and weather monitoring and measurement of pollution levels \cite{Butila2022}. These vehicles operate in diverse environments, such as urban areas, mountainous terrains, or in the wake of another vehicle, where unsteady aerodynamic effects are prevalent. In many cases, these atmospheric disturbances pose significant risks, as their spatial scale can be comparable to, or even larger than, the characteristic length scale of small aerial vehicles. 

Unsteady aerodynamic effects are commonly referred to as ``gusts", and manifest as flow disturbances of varying intensity and size, with the potential of posing great disruptions on flight performance and stability \cite{jones2022gust}. Much of this work focuses on vortex gusts, which typically originate in the wakes of wings, blades, or bluff bodies, and atmospheric turbulent processes. The strength and size of these disturbances are prescribed by two non-dimensional numbers, namely the gust ratio $G=u_g/u_\infty$ (where $u_g$ denotes the maximum tangential velocity of the vortex and $u_\infty$ represents the free-stream velocity), and radius $R_v$ \cite{jones2020taming}. Typically, the transition from moderate to extreme vortex dynamics is set at $|G|=1$ \cite{Fukami2023grasping}, at which the order of magnitude of the disturbances matches that of the free-stream, often leading to flow separation. In practice, a small aircraft may experience a wide range of gust-airfoil interactions, characterized by different gust ratios, sizes, and relative offsets between the airfoil and the gust. A deeper understanding of the individual effect of each parameter is crucial for improving the prediction of the aircraft's aerodynamic response and avoiding adverse outcomes. Here we review some literature that addresses the influence of the key airfoil and gust parameters considered in this work: gust ratio ($G$), gust size ($R_v$), vertical offset of the vortical gust ($y_0$), angle of attack ($\alpha$), airfoil thickness ($\tau$), and airfoil camber ($\eta$). 

Gust-airfoil interactions have been broadly characterized with emphasis on the influence of the gust ratio and the angle of attack. In general, larger aerodynamic responses have been observed for stronger gusts. This dependence was established early on by linear inviscid theories \cite{kussner1936,greenberg1947airfoil,sears1940operational}. It was later confirmed experimentally that these models yield accurate predictions and that the magnitude of the transients increases with $|G|$ \cite{corkery2018forces,AndreuAngulo2020gust}, even until gust impingement occurs in the regime of extreme aerodynamics ($|G|>1$) \cite{Jones2020unsteady}. While fluctuations of $|G|$ yield similar trends across flow regimes, specific responses change subtly depending on the flow conditions. For instance, \cite{weingaertner2020parallel} identified differences in the increment of the lift response, as well as in the location and extent of the separation bubble, with variations in $|G|$ between pre- and post-stall regimes for a NACA 0012 airfoil. Furthermore, as the magnitude of the aerodynamic transients increases with $G$, \cite{fukami2025re5000} reported the emergence of three-dimensional small-scale structures around $|G|\geq 4$, caused by spanwise instabilities. A similar trend is also observed in three-dimensional finite wings; however, the magnitude of the lift fluctuation is partially mitigated by gust-induced wingtip vortices, not observed in two-dimensional studies \cite{odaka2025:3dgust}. Moreover, the direction of the rotation of the vortex gust also has a significant influence on the evolution of the aerodynamic forces during a gust-airfoil interaction. In particular, for clockwise vortex gusts ($G>0$), the initial lift response is positive, while for counter-clockwise vortex gusts ($G<0$), the response is initially negative \cite{viswanath2010flapping}.

The influence of the angle of attack ($\alpha$) has been examined in several studies. In the context of gust-airfoil interactions, the notion of the effective angle of attack, defined by the sum of the geometric angle of attack and a gust-induced increment, is widely used. This interpretation is supported by the observation that the vortex gust introduces additional circulation around the airfoil, and the gust-induced local flow redirection near the leading edge effectively changes the location of the stagnation point, mirroring an increase in the geometric angle of attack. \cite{perrotta2017transverse} used this concept to predict the peak in the lift response and observed good agreement with experimental data, further noting that the effective angle of attack increases with the gust ratio. The same principle was also applied to successfully predict the flow response of an aeroelastic control surface to incoming transverse gusts in \cite{menon2020gusts}. Moreover, conditions similar to unsteady stall were reported in \cite{barnes2020angle} as a result of large gust-induced increases in the effective angle of attack for airfoils that presented no signs of stall in their unperturbed state. In addition, \cite{Engin2018} reported that the transient lift curves measured at different angles of attack were self-similar, and \cite{mmuriel2020lowRe} observed that the duration of the aerodynamic transients during a vortex-gust encounter was agnostic to $\alpha$.


The investigations regarding the influence of the vortex size suggest that a larger gust size expands the duration of the airfoil-gust interaction, and larger aerodynamic responses have been observed for larger gusts. For instance, the period of vortex-airfoil interaction was observed to grow with the size of the vortex core $R_v$ in \cite{mmuriel2020lowRe}. Regarding the magnitude of the lift peaks, vortex-gusts of increasing size were reported to yield larger aerodynamic fluctuations in \cite{barnes2020angle}. Moreover, the authors interpreted larger gust sizes as prolonged exposures to higher (effective) angles of attack, facilitating intense viscous interactions that enable the formation of a leading-edge vortex (LEV). Nonetheless, it was noted that the recovery period post-impingement mirrored those typical of smaller core sizes. 

The relative position of the vortex with respect to the airfoil, and in particular to the leading edge, has been identified as a relevant parameter as well. In general, past studies suggest that stronger aerodynamic responses are observed when the vertical offset between the vortex gust and the airfoil decreases. Based on the vortex trajectory, vortex-gust encounters are characterized as direct interactions, very close interactions, and close interactions \cite{Peng2014}. \cite{wilder1998parallel} and \cite{Barnes2018Re} reported that, for direct and very close interactions, the vortex gust is split between the upper and lower surfaces of the airfoil, disturbing both boundary layers and exacerbating the magnitude of the aerodynamic transients. On the other hand, the magnitude of the transients is mitigated when the distance between the airfoil and the gust increases. Furthermore, the investigations of \cite{horner1993asymmetric}, later confirmed by \cite{Peng2017asymmetric}, revealed an asymmetric distribution of the amplitude of the lift fluctuation against the vertical position of the vortex-gust with respect to the leading edge, while also highlighting the influence of the direction of the rotation of the vortex gust. 

The influence of the airfoil geometry has been addressed to some extent. Traditional studies have generally considered either flat plates or thin symmetric airfoils when characterizing the nonlinear dynamics observed during gust encounters. Existing work suggests that airfoil camber can affect the lift response when the gust has a streamwise component \cite{Young2020camber}, and in some cases attenuate the amplitude of the lift fluctuation \cite{Zhu2015camber}. \cite{Gementzopoulos2024blunt} investigated the effect of the geometry of the leading edge on the aerodynamic loads observed during transverse-gust encounters, and reported notable differences in the vorticity generation between the measurements obtained for sharp and blunt leading edges. Moreover, \cite{zhong2024geometrytransonic} examined the influence of airfoil geometry on the flow distributions observed in gusty transonic flows, and reported an attenuation of the lift response for thick airfoils. These investigations suggest that airfoil geometry may indeed influence the dynamics observed in vortex-gust encounters, suggesting an important direction for further research. 


The manuscript is structured as follows. A description of the dataset and approach used in this work is detailed in \S\ref{sec:methodology}. An outline of the main nonlinear vortex dynamics observed during a vortex-gust encounter is provided in \S\ref{sec:vortexDyn}. The analysis of the trends for each parameter of interest is presented in \S\ref{sec:results_indiv}, while \S\ref{sec:results_multi} discusses the combined effect of several gust and airfoil parameters. 
The principal findings of this study are summarized in \S\ref{sec:conclusions}. 

\section{Approach}
\label{sec:methodology}

\subsection{Computational setup}
\label{sec:ibpm}
We perform our numerical simulations at a fixed chord-based Reynolds number of $Re_c = u_\infty c/\nu=100$, where $u_\infty$ represents the free-stream velocity, $c$ is the chord length, and $\nu$ corresponds to the kinematic viscosity. 
Regarding the properties of the vortex gust, we explore the following (chord-based) parameters: 
angle of attack $\alpha$, gust ratio $G$, initial vertical gust position $y_0$, and gust radius $R_v$, which represents the distance from the core at which maximum tangential velocity is achieved. In terms of airfoil geometry parameters, we explore the effect of airfoil thickness $\tau$ and maximum camber $\eta$ on 4-digit NACA airfoils. All variables have been nondimensionalized such that the chord length $c=1$ and the free-stream velocity $u_\infty=1$. 

The vertical offset $y_0$ is defined with respect to the location of the leading edge. We consider angles of attack $0^\circ \leq \alpha \leq 20^{\circ}$, gust ratios $-2 \leq G \leq 2$, vertical offsets $-0.25 \leq y_0 \leq 0.25$, and gust radii $0.1 \leq R_v \leq 0.6$. To systematically explore variations in airfoil geometry, we examine a range of 4-digit NACA profiles with airfoil thicknesses $0 \leq \tau \leq 0.4$, where $\tau=0$ corresponds to a flat plate, and $\tau=1$ represents the extreme case in which the airfoil thickness matches the chord length. Regarding the airfoil camber, we consider maximum values $0.02 \leq \eta \leq 0.06$. The position of maximum camber $\xi$ is fixed at $\xi=0.4$. This position is chosen such that neither the leading edge nor the trailing-edge curvature is disproportionally accentuated, providing a general representation of a cambered airfoil. For all the cases considered here, the airfoil geometry, angle of attack, and position remain fixed during the gust encounter, and the analysis assumes two-dimensional incompressible flow conditions. 

\begin{figure}[t!]
\centering {
\vspace{-0.cm} 
{\hspace*{0cm}\includegraphics[width= 0.9\textwidth]{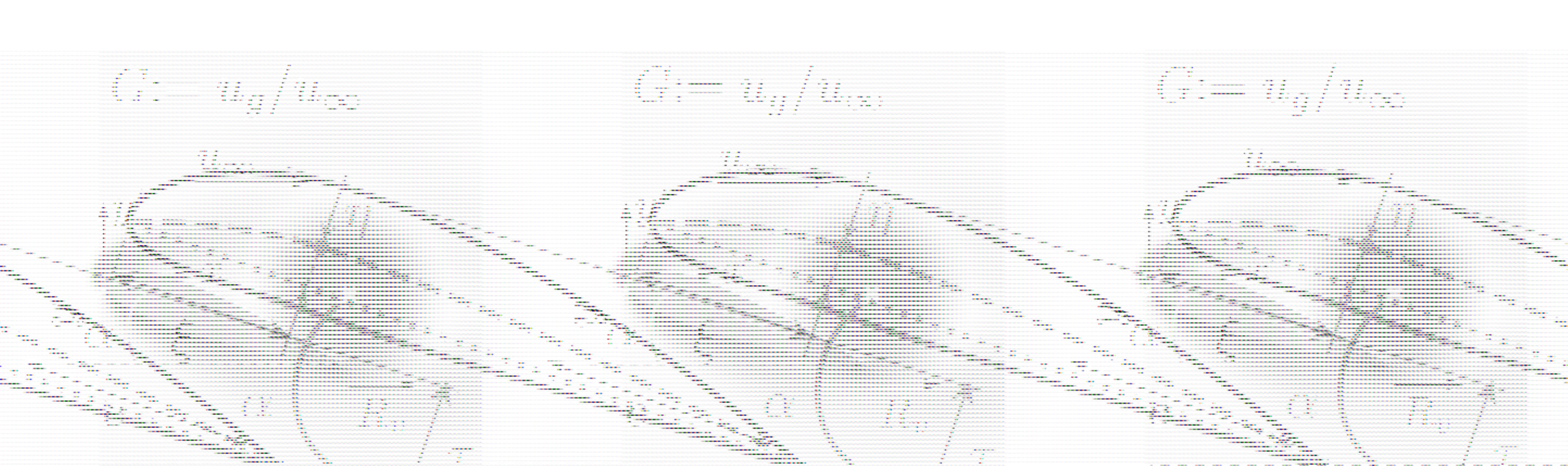}} }
\vspace{-0.2cm}
\caption{(Left) Diagram of the flow and airfoil variables included in the parameter space explored in this work. All variables of interest are highlighted in red. (Right) Schematic of the airfoil geometry parameters of 4-digit NACA profiles.
}
\vspace{-0.2cm}
\label{fig:airfoilSchematic}
\end{figure}

A schematic of the parameters that determine the different airfoil configurations considered in this work can be found in figure~\ref{fig:airfoilSchematic}. In particular, figure~\ref{fig:airfoilSchematic} (left) depicts a configuration that captures a vortical gust of radius $R_v$, and strength $G$, as it approaches an airfoil of chord-length $c$ and at an angle of attack $\alpha$. According to the sign convention adopted in this work, a vortex gust characterized by a counter-clockwise rotation is considered to have a positive orientation, and is therefore indicated by a positive value of $G$. We define the gust vortex in terms of vorticity with 
\begin{equation}
    \label{eq:vortex}
    \omega_r(r) = \frac{u_g}{R_v}\left(2-\frac{r^2}{R_v^2}\right)\exp{\left( \frac{1-r^2}{2R_v^2}\right)},
\end{equation}
where $r$ represents the distance with respect to the vortex core \cite{taylor1918dissipation}. Traditionally, the gust ratio $G$ is defined as $G=u_g / u_\infty$, that is, the relationship between the peak gust tangential velocity $u_g$ and the free-stream velocity $u_\infty$. A diagram is provided in figure~\ref{fig:airfoilSchematic} (right) to depict the two variables of interest that relate exclusively to the airfoil geometry, namely: airfoil thickness $\tau$ and maximum camber $\eta$.

The immersed boundary projection method is used to simulate the flow \cite{taira2007IMBP,colonius2008imbp}. The total extent of the computational domain is $L_x=72$ in the $x$-direction and $L_y=40$ in the $y$-direction. The multi-grid is formed by a total of 5 subdomains, the finest one spanning between $-4 \leq x \leq 5$ and $-2 \leq y \leq 3$ with $M=1152$ and $N=640$ cells, respectively. To ensure consistency across different gust encounters, the origin of our reference henceforth coincides with the leading edge (see figure~\ref{fig:airfoilSchematic}). The time-step is chosen such that the CFL number is 0.256 for all simulations. To ensure convergence, all baselines included in this dataset (\textit{i.e.}~all initial conditions preceding the vortex-gust encounter) were obtained after $t=40$ convective time units since initialization. To preserve uniformity, all vortical gusts are introduced at the same temporal instance $t_{0}=40$, and at an initial location along the $x$-axis corresponding to $x_0=-2$ with respect to the leading edge. Further details regarding the convergence studies performed for this mesh are given in appendix \ref{app:convergence}.

\subsection{Force element analysis}
\label{sec:forceElement}
To interpret the vorticity fields during a gust encounter, we incorporate the force-element analysis \cite{chang1992potential,Zhang2015force}. Let us consider the velocity potential $\phi$, such that $\nabla \phi = \bsu$, where $\bsu = \bsu(\bsx,t)$ is the instantaneous velocity field. This potential is chosen to satisfy the geometric condition $\bsn \cdot \nabla \phi_i = -\bsn \cdot \bse_i$, where the right-hand side represents the projection of the surface normal onto the $i$-th basis vector of the orthonormal reference frame $\{ \bse_1, \bse_2, \bse_3 \}$ spanning $\mathbb{R}^3$. The use of this auxiliary velocity potential, along with the implementation of the divergence theorem on the velocity field (assumed to be solenoidal), allows us to extract the projection of the body and pressure forces onto each element $\bse_i$ as
\begin{equation}
\label{eq:forceElement}
    F_i = -\int_V \bsu \times \bsomega \cdot \nabla\phi_i \text{d}V  +\frac{1}{Re} \int_S \bsn \times \bsomega \cdot (\nabla\phi_i +\bse_i)\text{d}S,
\end{equation}
where the first term has been rewritten as a volume integral over $V$ involving the Lamb vector \cite{wu2006vorticity}. Given that this work is concerned with the impact of the transient vortical structures observed during a vortex-gust encounter, we restrict our focus to the first term in~\eqref{eq:forceElement}, and henceforth refer to the terms $f_{D}=-(\bsu \times \bsomega) \cdot \nabla \phi_x$ and $f_{L}=-(\bsu \times \bsomega) \cdot \nabla \phi_y$ as drag- and lift-force elements, respectively. 

\subsection{Vorticity production analysis}
\label{sec:vortProd}
To further characterize the mechanisms underpinning the generation and evolution of strong vortical structures during a gust encounter, we examine the source of vorticity. Specifically, we consider the vorticity production flux following the definition first provided in \cite{lighthill1963introduction}, and revisited in \cite{morton1984vorticity}, \cite{hornung1989vorticity} and \cite{panton2013incompressible}. Let us define the vorticity flux tensor
\begin{equation}
    \label{eq:J}
    \bsJ \equiv -\nu \nabla \bsomega.
\end{equation}
For two-dimensional flows, the vorticity transport equation is 
\begin{equation}
    \label{eq:vortTransport}
    \frac{D\bsomega}{Dt} = \nu \nabla^2 \bsomega,
\end{equation}
where the right-hand side represents the rate of viscous diffusion of vorticity. To better characterize the vorticity generation along the airfoil surface, we adopt a local curvilinear coordinate where $(\bss,\bsn)$ correspond to the wall-normal and surface unit vectors, respectively, at each location along the airfoil surface. In this reference, the instantaneous velocity field is expressed as $\bsu(\bsx,t) = [u_n(s,n,t),u_s(s,n,t)]$. 

We are particularly interested in the component of the vorticity flux along the wall-normal direction and evaluated at the wall ($n=0$), given by
\begin{equation}
    \label{eq:Jn0}
    J_{n,0}=\left(\bsn\cdot\bsJ\right)_{0}=-\nu \left(\bsn\cdot \nabla \bsomega\right)_{0},
\end{equation}
which quantifies the generation of vorticity with the no-slip condition. In this reference, the vorticity can be written as
\begin{align}
\label{eq:vortCurv}
    \omega &= \frac{\partial u_s}{\partial n} - \frac{\partial u_n}{\partial s} + \frac{u_s}{r} \notag \\
    &= \frac{\partial u_s}{\partial n} - \frac{\partial u_n}{\partial s} + \kappa u_s,
\end{align}
where $\kappa=1/r$ represents the curvature of the airfoil surface. Evaluating the projection of the vorticity gradient onto the wall-normal direction $\bsn$ gives
\begin{equation}
    \label{eq:domegadn0}
    \left(\frac{\partial \omega}{\partial n}\right)_{n=0}=\left(\frac{\partial^2 u_s}{\partial n^2} +\frac{1}{r}\frac{\partial u_s}{\partial n}\right)_{n=0},
\end{equation}
at the airfoil surface. Although not explicitly included in the vorticity transport equation, surface pressure plays a fundamental role in the process of vorticity generation at solid boundaries. The connection between the wall-normal component of the vorticity gradient and pressure is established through the momentum equation. In particular, the momentum equation in the $s-$direction, evaluated at the wall, is reduced to 
\begin{equation}
    \label{eq:smomentum}
    \left(\frac{\partial u_s}{\partial t}+\frac{1}{\rho}\frac{\partial p}{\partial s}\right)_{n=0}=\nu \left(\frac{\partial^2 u_s}{\partial n^2} +\frac{1}{r}\frac{\partial u_s}{\partial n}\right)_{n=0},
\end{equation}
where $\rho$ represents the fluid density, body forces are excluded from the analysis, and the no-slip condition is enforced at the airfoil surface. According to \eqref{eq:smomentum}, for fluid elements at the wall ($n=0$), vorticity-producing unbalanced shear-stresses are balanced by both the pressure gradient and wall motion \cite{lighthill1963introduction,morton1984vorticity}. 

The wall-normal component of the vorticity flux in \eqref{eq:Jn0} can now be written as
\begin{align}
\label{eq:Jn0_final}
    J_{n,0} &=-\left(\frac{\partial u_s}{\partial t}+\frac{1}{\rho}\frac{\partial p}{\partial s}\right)_{n=0}, 
\end{align}
where the wall-normal component of the vorticity gradient in \eqref{eq:domegadn0} has been replaced by the left-hand side of \eqref{eq:smomentum}. Herein, we define the quantity
\begin{equation}
    \label{eq:curv}
    J_{n,0}^c=\nu\left(\frac{1}{r}\frac{\partial u_s}{\partial n}\right)_{n=0},
\end{equation}
as the contribution of the curvature term to the total wall-normal vorticity flux. This analysis is particularly meaningful in the context of gust encounters, as it provides a direct connection between the gust-induced changes in the surface pressure and modified vorticity at the airfoil surface. These changes in the vorticity production levels affect the net circulation around the airfoil, and therefore influence the temporal evolution of the lift during a gust encounter. 

\subsection{Baseline flows in our dataset}
\label{sec:baselines}
This subsection provides a brief overview of the baseline (unperturbed) state of the airfoil wakes. At this $Re$, all baseline flows are at steady states with no vortex shedding. The angle of attack $\alpha$ and the airfoil parameters $(\tau,\eta)$, however, determine the initial properties of the flow around the airfoil before the gust-airfoil interaction begins. We characterize the baseline states through the lift $C_L=L/(\frac{1}{2}\rho u_\infty^2)$ and drag $C_D=D/(\frac{1}{2}\rho u_\infty^2)$ coefficients, where $L$ and $D$ denote the lift and drag forces experienced by the airfoil, respectively. We consider three symmetric and two cambered airfoils. The chord-wise location of maximum camber is fixed at $\xi=0.4$ to highlight the influence of the camber.

\begin{figure}[t!]
\centering {
\vspace{-0.0cm}
{\hspace*{0cm}\includegraphics[width= 1\textwidth]{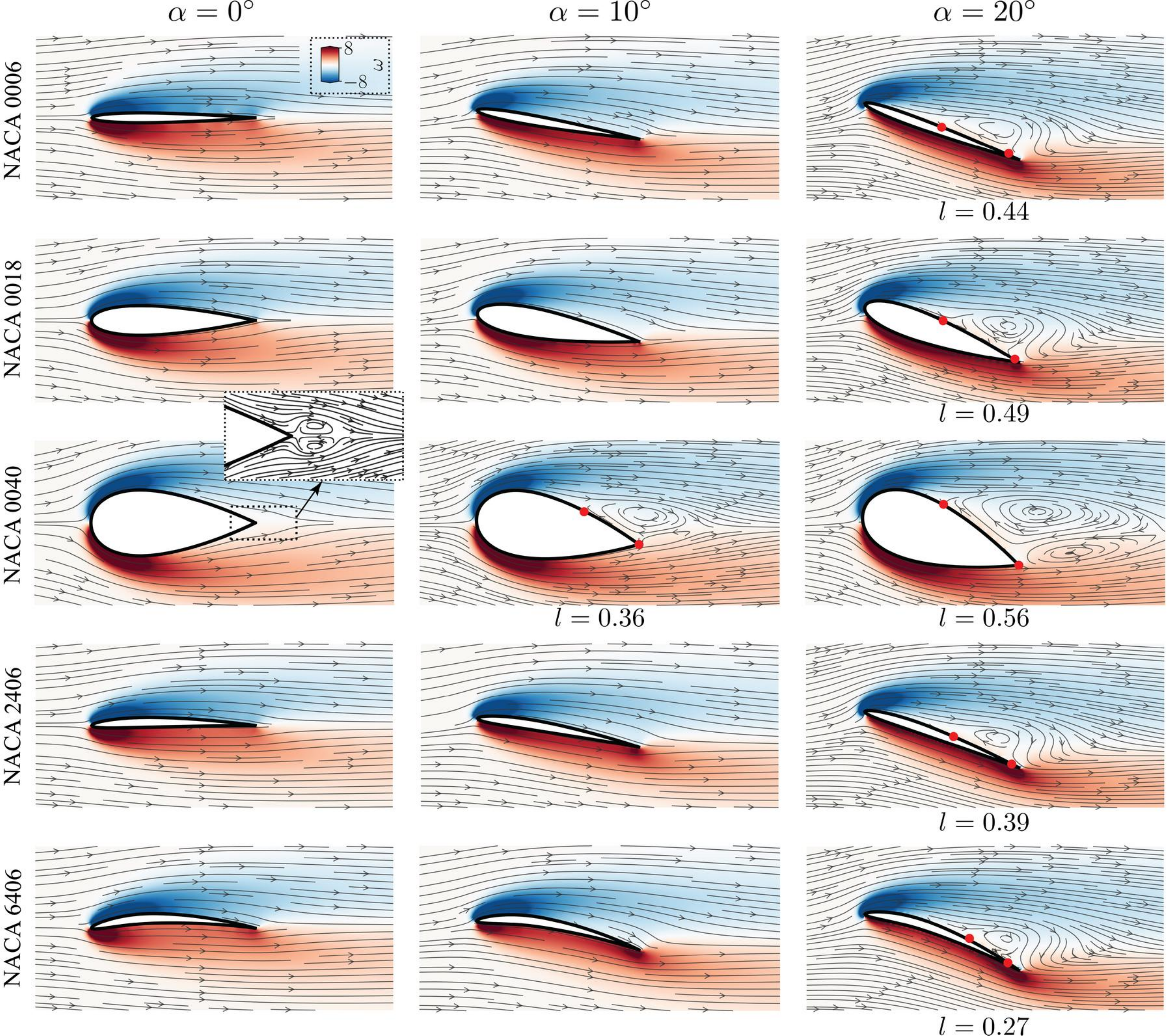}} }
\vspace{-0.5cm}
\caption{Baseline streamlines and vorticity (colored contours) fields of three symmetric airfoils at different angles of attack $\alpha$. Red markers indicate the bounds of the separated region, and its chordwise extent is denoted by $l$. 
}
\vspace{-0.cm}
\label{fig:baselines_vort}
\end{figure}

\begin{figure}[t!]
\centering {
\vspace{-0.cm}
{\hspace*{0cm}\includegraphics[width= 0.9\textwidth]{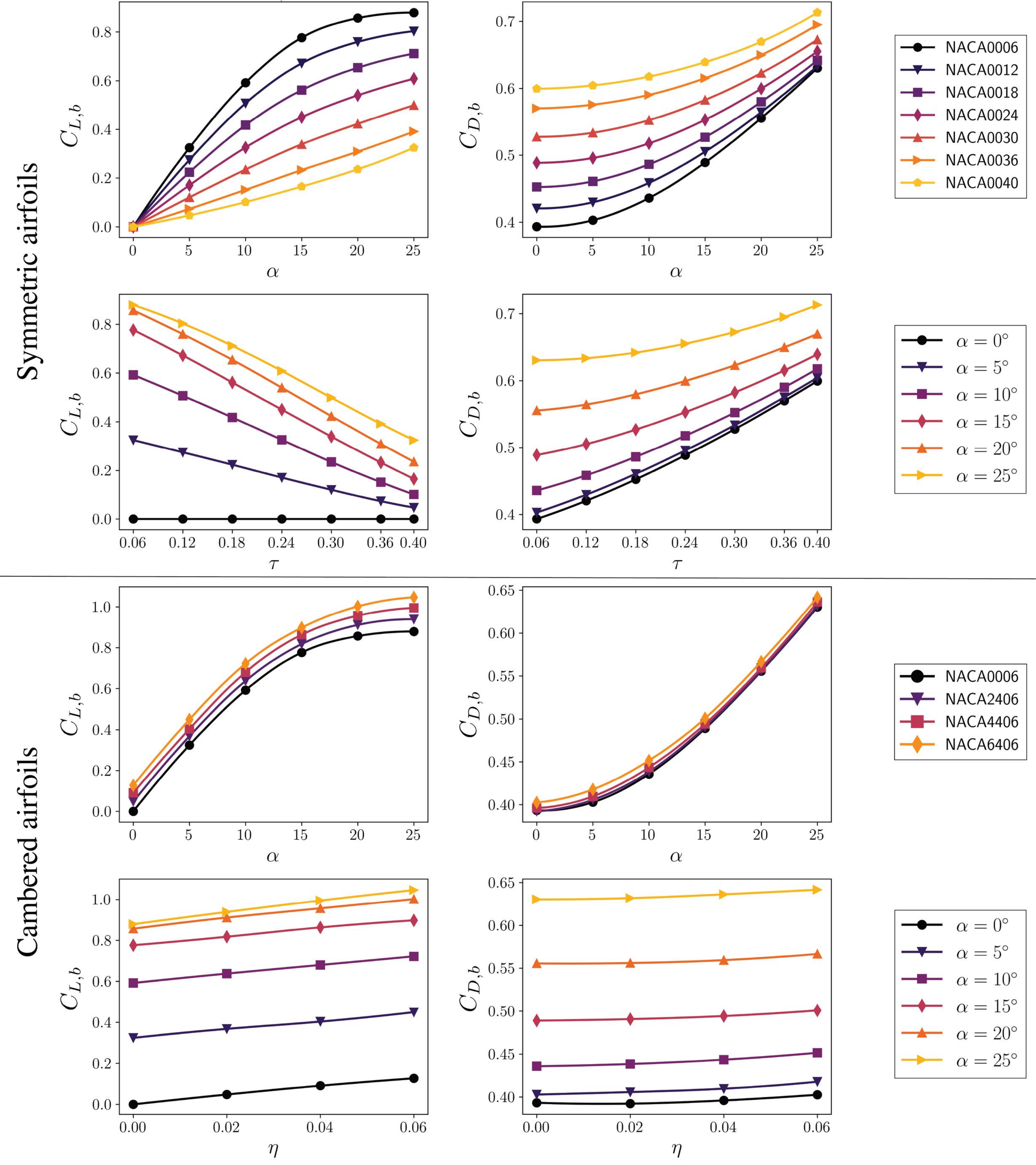}} }
\vspace{-0.cm} 
\caption{Baseline lift $C_{L,b}$ (first column) and drag $C_{D,b}$ (second column) coefficients of symmetric and cambered 4-digit NACA profiles at different angles of attack $\alpha$. 
}
\vspace{-0.cm}
\label{fig:baselines}
\end{figure}

Let us first discuss the influence of $\alpha$ and airfoil geometry on the baseline flows shown in figure \ref{fig:baselines_vort}, in which the velocity fields are shown with streamlines, and the vorticity fields are visualized with colored contours. Flow separation originates near the trailing edge and develops upstream in configurations with increasing $\alpha$, regardless of the airfoil geometry. The extent of the separated region is denoted as $l$ and represents the projection of this region onto the chord-wise direction (for which the bounds are determined as the locations at which $\pp u_s/\pp n$ changes its sign). Notably, while an intermediate angle of attack of $\alpha=10^\circ$ does not lead to a separated region for $\tau\leq0.18$, regardless of the camber, the extent of the separation reaches a value of $l=0.36$ for $\tau=0.4$. However, this length remains comparable across symmetric airfoils at higher angles of attack $\alpha=20^\circ$, and decreases for cambered airfoils. The influence of the separated region is reflected in the lift $C_{L,b}$ (left) and drag $C_{D,b}$ (right) coefficients for the five baseline states at different angles of attack shown in figure~\ref{fig:baselines}. The first rows correspond to symmetric airfoils, while the third and fourth rows correspond to cambered airfoils. 

Baseline lift increases with $\alpha$ for all airfoil geometries, and flattens at high angles of attack \cite{abbott1959theory,leishman2006principles}. This trend is directly related to the presence of flow separation at high angles of attack observed in figure~\ref{fig:baselines}. For symmetric airfoils, thick airfoils yield less lift than their thin counterparts, although the flattening of the lift curve occurs at lower angles of attack for thin geometries. For cambered airfoils, the lift increases with a slope independent of the airfoil geometry, and the flattening of the curves with increasing $\alpha$ occurs similarly for all airfoils. We observe a steady decrease in lift, with a steeper slope for larger $\alpha$, with increasing airfoil thickness $\tau$. On the other hand, the lift increases with $\eta$ with a slope independent of $\alpha$ for all cambered airfoils. 

Baseline drag, on the other hand, increases with $\alpha$ for all geometries, coinciding with the growth of the separated region that emerges at high incidences. We observe a steeper rate of growth on the drag curves for thin airfoils, and a less dramatic increase for thick airfoils. This coincides with the slower rate of growth of the length of the separated region observed for thick geometries. Moreover, the baseline drag also increases with $\tau$, for which the separated region originates at lower incidences. In contrast, the slope of the baseline drag against $\alpha$ is similar across the cambered geometries, and increases for cambered airfoils across all incidences.

\section{Description of vortex dynamics during a vortex-gust encounter}
\label{sec:vortexDyn}
\begin{figure}[t!]
\centering {
\vspace{0.05cm}
{\hspace*{-0.15cm}\includegraphics[angle=0, width= 1\textwidth]{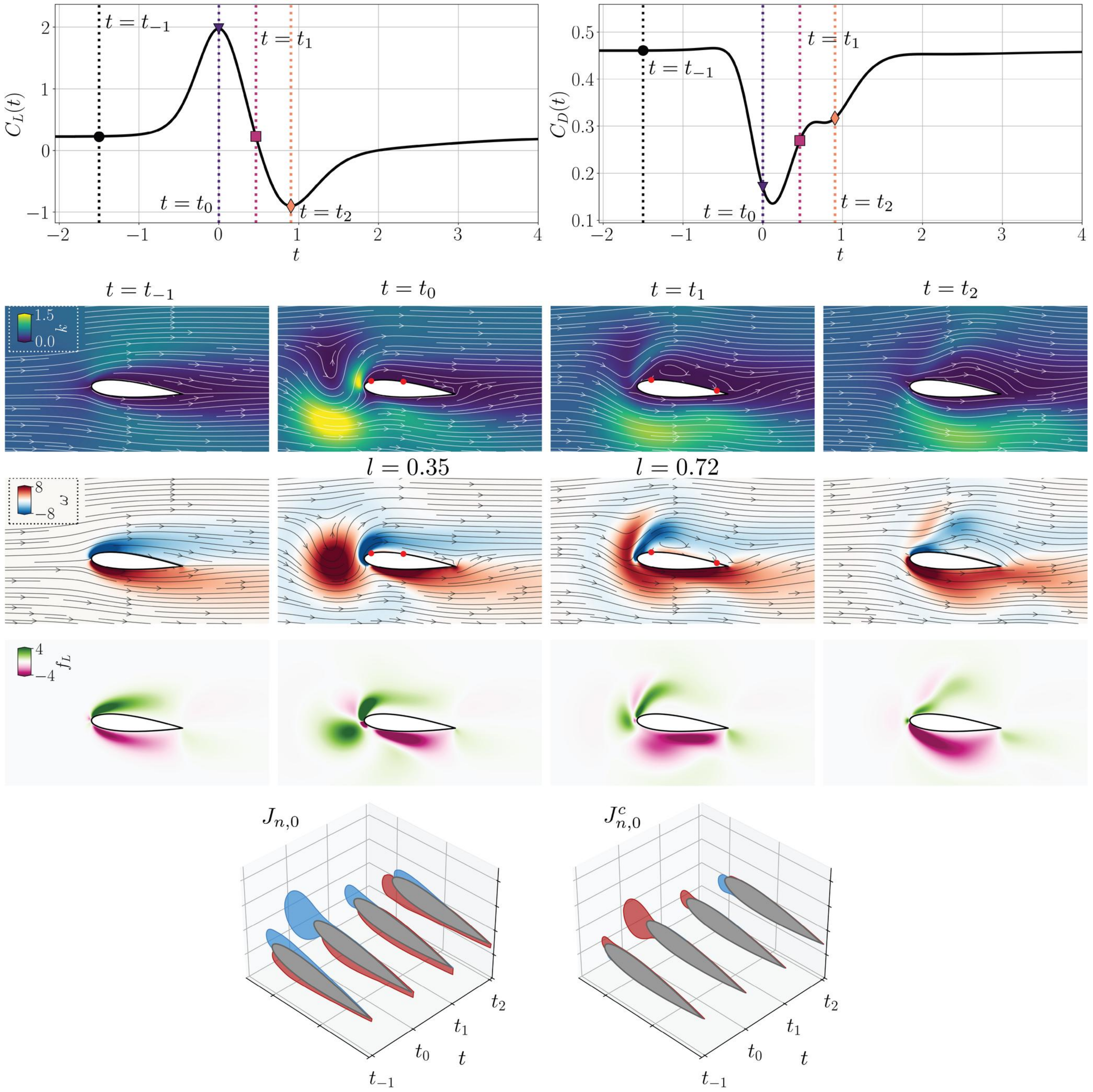}} }
\vspace{-0.4cm}
\caption{Lift $C_L(t)$ and drag $C_D(t)$ coefficients, kinetic energy $k$, velocity $(u,v)$, vorticity $\omega$, and lift-element $f_L$ fields, along with vorticity production fluxes ($J_{n,0},J_{n,0}^c$), observed during a during a vortex gust encounter by a NACA 0018 and $(G,R_v,y_0,\alpha)=(2,0.25,-0.1,5^\circ)$. 
}
\vspace{-0.3cm}
\label{fig:vortProd_intro}
\end{figure}

Let us provide a general overview of the evolution of the flow physics during a vortex-gust encounter. First, we discuss a canonical case presented in figure~\ref{fig:vortProd_intro} for a NACA 0018 with $(G,R_v,y_0,\alpha)=(2,0.25,-0.1,5^\circ)$. These values represent moderate choices, particularly for $\alpha$ and $\tau$, within the explored parameter ranges, selected to depict a typical vortex-gust interaction while avoiding extreme angles of attack or large thicknesses that could produce highly specific flow responses. Moreover, this case with $y_0=-0.1$ was selected to showcase a close interaction, in which the coherence of the vortex gust is greatly distorted post-impingement and depicts a much richer set of nonlinear dynamics with respect to other interactions with larger vertical offsets. This description will serve as a benchmark for later investigations into the influence of individual airfoil and gust parameters in \S\ref{sec:results_indiv}.

The curves at the top of figure~\ref{fig:vortProd_intro} show the evolution of lift $C_L(t)$ and drag $C_D(t)$ coefficients. Four specific time instances, denoted as $t=t_i$, are highlighted on these curves to provide a characterization of the state of the system throughout the interaction. The velocity fields ($u,v$) (streamlines), kinetic energy distributions $k=\frac{1}{2}(u^2+v^2)$ (colored contours), vorticity fields $\omega$, and lift-elements $f_L$ corresponding to each time instance are shown in the middle plots in figure~\ref{fig:vortProd_intro}. The vorticity production flux along the airfoil surface $J_{n,0}$, as well as the contribution of the curvature term $J_{n,0}^c$, are presented in the bottom subfigures in figure~\ref{fig:vortProd_intro}, in which red and blue shades denote regions of positive and negative values, respectively. 

The first temporal instance, $t=t_{-1}$, is presented as the initial undisturbed state, in which the flow remains fully attached to the airfoil. The kinetic energy distribution exhibits low values around the airfoil as a result of the no-slip condition, and higher values in the freestream. The lift-element field highlights positive and negative lift contributions along the upper and bottom surfaces, respectively, with a net positive lift resulting from the greater contribution of the positive element on the upper surface. Moreover, we observe a region of negative wall-normal vorticity production above the stagnation point, and a positive region below the stagnation point. Both distributions become more intense near the leading edge, mainly due to the enhanced contribution of the curvature term $J_{n,0}^c$. The exacerbated effect of this term near the leading edge seems intuitive, considering that the curvature $\kappa=1/r$ becomes increasingly large in this region, significantly boosting the local vorticity production levels (see second term in \ref{eq:smomentum}). The surface pressure gradients are also heightened near the leading edge, resulting in a net positive lift due to the asymmetric surface pressure distribution between the lower and upper surfaces. 

As the vortex gust approaches the airfoil, it redirects the streamlines around the airfoil, affecting the pressure distribution, net circulation, and therefore the aerodynamic loads. For $G>0$, negative vorticity starts accumulating at the leading edge, causing local flow acceleration around the leading edge. In the instance in which the vortex core impinges on the leading edge $(t=t_0)$, the stagnation point has shifted aft, and flow acceleration occurs at the leading edge (see corresponding $k$ field). We observe a laminar separation bubble (LSB) near the leading edge on the upper surface, with $l=0.35$, as a result of the heightened pressure gradient in this region. This enhancement of the surface pressure gradients is signaled by an increase in the vorticity production levels around the leading edge, and in particular the contribution of the curvature term $J_{n,0}^c$, driven by the boosted velocity gradients in this region. The accumulation of new vorticity near the leading edge corresponds to a positive lift-element $f_L$, contributing to a positive increase in lift. As the gust approaches the airfoil, the flow traveling from the stagnation point and around the leading edge experiences great acceleration, and it is indeed opposing the direction of the free stream. We interpret this enhanced acceleration as a negative contribution toward drag, as it alleviates total shearing stresses. Furthermore, we identify a slight time lag between the first lift peak and the time instance of minimum drag. Notably, we observe a small recirculation region near the trailing edge that does not have a relevant impact on the net lift force, as revealed by lift-element analysis.   

Upon impingement, the mass of the vortex gust is split into two portions: above and below the leading edge. We will henceforth refer to them as the upper and lower portions of the gust. The newly generated gust-induced vorticity at the leading edge begins to roll up into a vortical structure above the upper surface. The strong vorticity ingestion of the lower portion of the gust results in a thickened boundary layer. The acceleration provided by the lower portion of the gust on the boundary layer leads to the generation of additional positive vorticity, coinciding with a negative surface pressure gradient. Meanwhile, vortex roll-up occurs on the upper surface (as a result of pressure gradient alleviation at the leading edge), subtracting circulation from the boundary layer and yielding a lower lift force. The LSB that emerged pre-impingement shifts and grows aft, reaching a value of $l=0.72$ at $t=t_1$ (time instance at which $C_L=C_{L,b}$). At this point, the stagnation point has moved upstream, and the pressure gradient at the leading edge has subsided close to the baseline value. 

Over time, the lower portion of the gust continues to thicken and accelerate the boundary layer on the bottom surface, further enhancing the positive vorticity production levels. As a result of the continued introduction of positive vorticity from the lower portion of the gust, the net circulation around the airfoil becomes negative. This trend subsides after the second lift peak ($t=t_2$), which marks the instance at which the lower portion of the gust ceases to provide additional acceleration and positive vorticity to the boundary layer. This instance also defines the onset of lift recovery. On the upper surface, however, the upper portion of the gust and the rolled-up vortical structure have formed a dipole (vortex pair), and the LSB has steadily contracted to the point of full reattachment. From this point forward, the system will gradually return to its original unperturbed (baseline) state. 

\section{Characterizing the effect of individual parameters on the nonlinear dynamics observed in gust encounters}
\label{sec:results_indiv}
This section examines the changes observed from the variation of one parameter at a time, while the others remain fixed. Each section is self-contained and may be read in any order. Studies regarding the combination of several parameters are compiled in \S\ref{sec:results_multi}.

\subsection{Effect of gust ratio}
\label{sec:gust}
\begin{figure}[t!]
\centering {
\vspace{0.05cm}
{\hspace*{-0.cm}\includegraphics[angle=0, width= 0.95\textwidth]{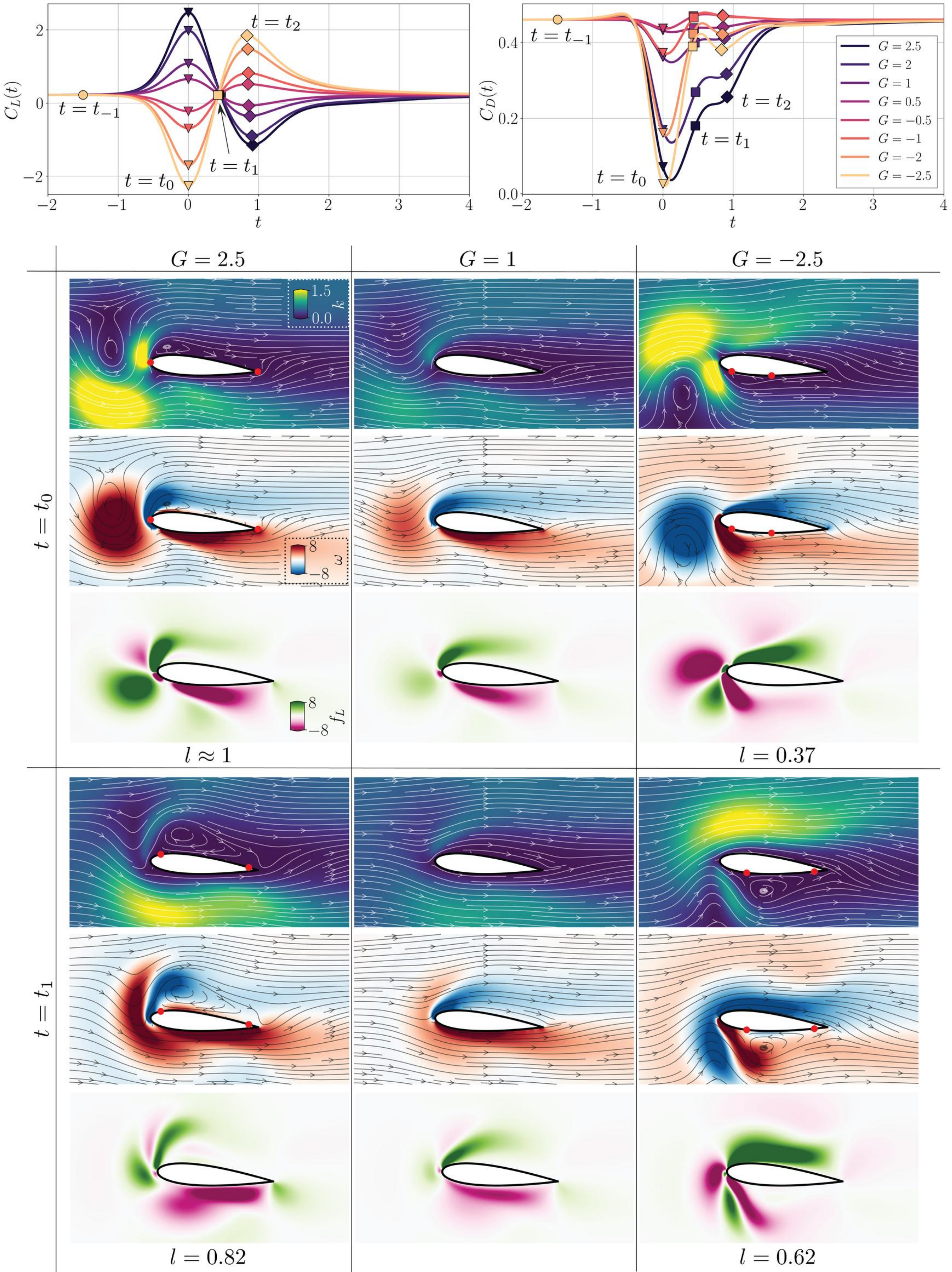}} }
\vspace{-0.cm} 
\caption{Influence of $G$ on $C_L(t)$ and $C_D(t)$, kinetic energy $k$, velocity (streamlines), vorticity $\omega$, and lift-element $f_L$ fields, observed during a vortex gust encounter by a NACA 0018 and $(R_v,y_0,\alpha)=(0.25,-0.1,5^\circ)$.  
}
\vspace{-0.cm} 
\label{fig:results_G}
\end{figure}

The effect of the gust ratio $G$ on the behavior of the flow around the airfoil during a gust encounter is determined by both the direction and strength of the vortex gust. Let us consider an example of a NACA 0018 profile with $(R_v,y_0,\alpha)=(0.25,-0.1,5^\circ)$. The main results of this analysis are presented in figure~\ref{fig:results_G}. 

Two main observations are drawn from the lift curves: the amplitude of the lift response $|\Delta{C}_{L}|=|\max{(C_L)}-\min{(C_L)}|$ (or lift fluctuation) increases with $|G|$, and the sign of the first lift peak matches that of $G$ (\textit{i.e.}~we observe a first positive peak for $G>0$, and a first negative peak for $G<0$). As the vortex gust approaches the airfoil, an increased value of $|G|$ will lead to an accentuated deviation of the streamlines and a stronger local flow acceleration near the leading edge (see velocity and kinetic energy fields for $G=\{1,2\}$ in figure~\ref{fig:results_G}). As a result, the vorticity production flux and the pressure gradient at the leading edge will be enhanced, amplifying the net circulation around the airfoil, and therefore yielding a stronger lift response. At this $Re$, 
the flow remains fully attached for $G=1$. We also report an increase in the extent of the LSB that emerges near the leading edge for $G=2.5$ ($l\approx 1$) against the case shown in figure~\ref{fig:vortProd_intro} for $G=2$. Moreover, for $G=2.5$, we observe a region of strong flow reversal coming from the bottom surface, bending around the trailing edge, and moving upstream along the upper surface that reaches the end of the LSB. The separated region that emerges from the trailing edge is reminiscent of near-stall conditions, and a reflection of the considerable magnitude of the vortical disturbance. In this case, we report only the total extent of the separated region $l$, since it includes both the LSB and the trailing-edge recirculation zone. 

Let us turn our attention to the case in which $G<0$. As the gust approaches the airfoil, downward local momentum is induced at the leading edge, shifting the stagnation point to the upper surface. This flow redistribution leads to the establishment of a reversed suction peak at the leading edge, providing a negative contribution towards lift. In this case, we observe a region of high kinetic energy above the leading edge, and an LSB near the leading edge on the lower surface of extent $l=0.37$. 

We observe a negative trend in the transient drag for all values of $G$, and the differences in the drag curves between cases of equal $|G|$ are attributed to the asymmetries introduced by the non-zero angle of attack $\alpha$ and the offset in the initial vertical position of the gust $y_0$ with respect to the leading edge. We attribute the negative trend for all gust ratios, regardless of the direction of the vortex gust, to the acceleration of the flow near the leading edge. At low incidences, the gust-induced increment in the effective angle of attack appears to be larger in magnitude than the geometric effective angle of attack, since the amplitude of the drag fluctuations is similar between curves of the same $|G|$. Though the direction of the accelerated flow at the leading edge is dictated by the sign of the gust ratio, it provides local alleviation of the shear stresses in this region, leading to a negative drag fluctuation. 

The influence of $G$ follows the same trends post-impingement shown in figure~\ref{fig:results_G}. 
For $G>0$, the lower portion of the gust leads to a more prominent thickening of the boundary layer and a more acute acceleration of the flow on the lower surface for larger gust ratios. As a consequence, the positive lift element on the lower surface, and therefore the secondary lift peak, is also more pronounced for higher values of $|G|$. At $t=t_1$, the extent of the separated region has increased to $l=0.82$, which is also larger than the separated region observed for $G=2$ in figure~\ref{fig:vortProd_intro}. Notably, we do not report a separated region at this stage for $G=1$. Moreover, we identify the formation of a coherent leading-edge vortex (LEV) \cite{eldredge2019LEV} for $G=2.5$, as a result of the transient near-stall conditions induced by the gust. The post-impingement dynamics for $G=-2.5$ lead to the expansion of the LSB to a length $l=0.62$ on the lower surface. Moreover, in this case, the upper portion of the vortex gust is responsible for the thickening of the boundary layer on the upper surface, coinciding with a positive lift element and yielding a positive secondary lift peak.  

The principal trends observed in this section agree with previously established findings: the magnitude of the lift fluctuations increases with $G$ \cite{kussner1936,corkery2018forces}, and the direction of rotation of the vortex gust determines the sign of the first lift peak \cite{viswanath2010flapping}. Nonetheless, we attribute an increase in the magnitude of the lift fluctuations to accentuated vorticity production levels at the leading edge, signaled by enhanced lift elements in this region. Moreover, we report an increase in the magnitude of the drag fluctuation with $|G|$.

\subsection{Effect of size of the vortical gust}
\label{sec:size}

\begin{figure}[t!]
\centering {
\vspace{-0.02cm}
{\hspace*{-0.cm}\includegraphics[angle=0, width= 0.9\textwidth]{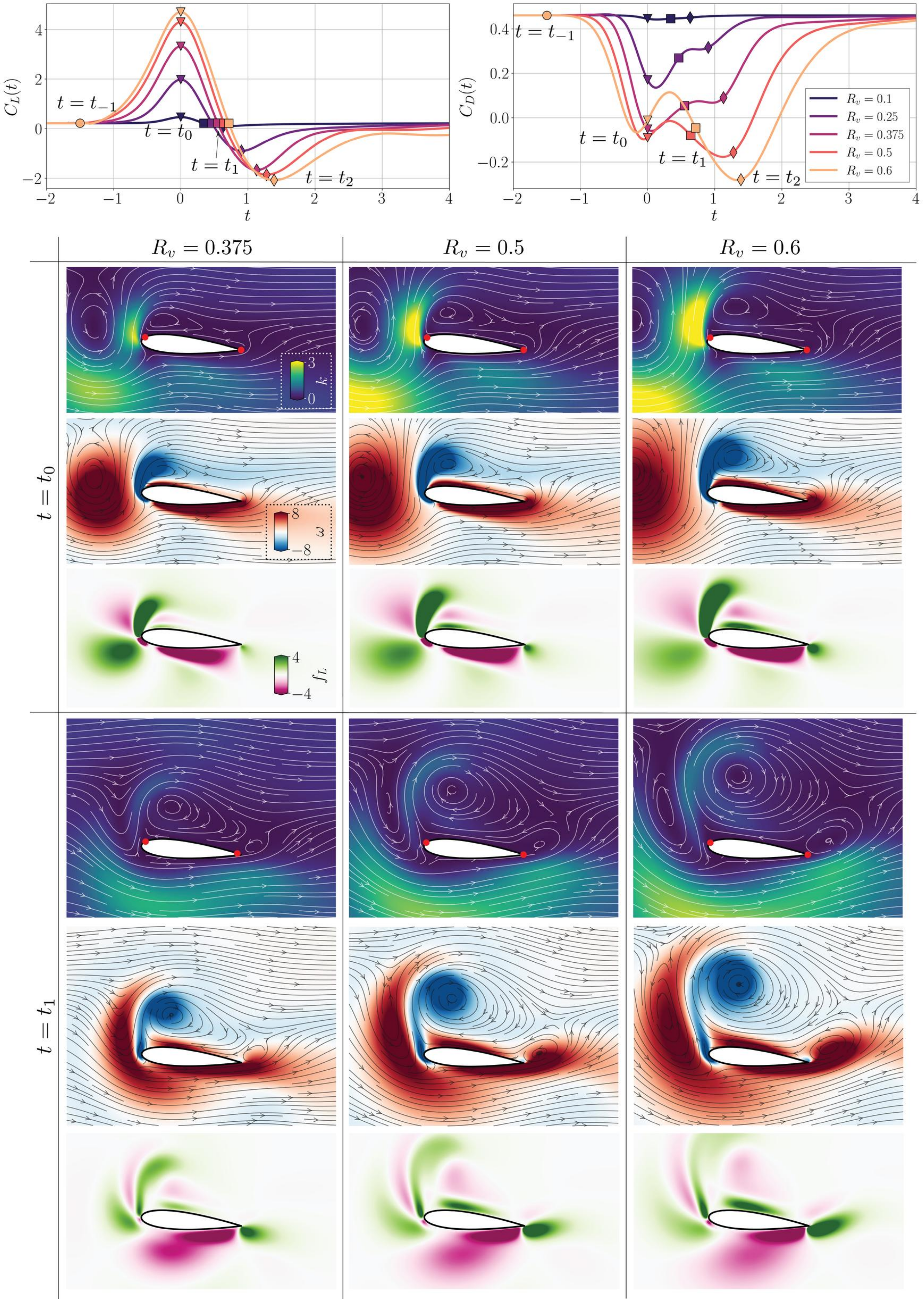}} }
\vspace{-0.cm} 
\caption{Influence of $R_v$ on $C_L(t)$ and $C_D(t)$, kinetic energy $k$, velocity (streamlines), vorticity $\omega$, and lift-element $f_L$ fields, observed during a vortex gust encounter by a NACA 0018 and $(G,y_0,\alpha)=(2,-0.1,5^\circ)$.  
}
\vspace{-0.cm}
\label{fig:results_rv}
\end{figure}

Let us now turn our attention to the influence of the vortex size on the dynamics observed during a vortex-gust encounter. To illustrate the effect of this parameter, we consider the case with a NACA 0018 profile with $(G,y_0,\alpha)=(2,-0.1,5^\circ)$. The results of the analysis on the influence of $R_v$ can be observed in figure~\ref{fig:results_rv}. In this analysis, the net circulation of the vortex remains the same with $R_v$, and the vorticity gradient in the radial direction varies with $R_v$. 

The effect of the gust radius $R_v$ on the lift and drag curves is obvious: the amplitude of both lift and drag fluctuations is accentuated by larger gust sizes. As $R_v$ increases, 
its effective area of influence on the surrounding flow expands, as indicated by the streamlines and kinetic energy distributions in figure~\ref{fig:results_rv}. Moreover, larger gust sizes correspond to longer interactions, in which vorticity strongly accumulates at the leading edge for longer time periods \cite{barnes2020angle}. This naturally translates as stronger vortical structures of increasing strength and spatial extent, as revealed by the vorticity and lift-element distributions, and the enhanced kinetic energy at the leading edge. Notably, as the gust size increases, flow separation emerges from the trailing edge and its extent grows along the upper surface. While for $R_v=0.25$ only an LSB was reported on the upper surface, we observe both an LSB and a massive separated region on the upper surface for $R_v>0.25$. Indeed, for larger gust sizes, $R_v=\{0.5,0.6 \}$, we observe full separation on the upper surface ($l\approx 1$). Moreover, greater recovery periods are attributed to larger gust sizes, as reported in \cite{barnes2020angle}. 

All configurations showcased in figure~\ref{fig:results_rv} exhibit a decreased drag upon impingement ($t=t_0$) with respect to the undisturbed case ($t=t_{-1}$). Notably, the magnitude of the lift variation at $t=t_0$ increases significantly from $R_v=0.1$ to $R_v=0.375$, after which it flattens. We attribute this trend to the similarity across the extent of the recirculation region on the upper surface for all cases $R_v\geq 0.375$. Note that $l\approx 1$ for all configurations and time instances showcased in figure~\ref{fig:results_rv}.

We observe a distinct LEV of increasing size and coherence for larger gust sizes. We attribute this to a prolonged time window of vorticity generation and accumulation at the leading edge for large $R_v$, which precipitates the formation of a stronger LEV via vortex roll-up upon impingement. At this stage, the entirety of the upper surface is part of the recirculation region, as shown by the streamlines in figure~\ref{fig:results_rv}. Moreover, we observe a secondary vortical structure at the trailing edge that becomes more prominent as $R_v$ increases. This structure has a positive contribution toward lift, as revealed by force-element analysis, which presumably mitigates the amplitude of the secondary lift peak. 

The growing size and strength of the vortical structures identified above the upper surface at $t=t_1$ lead to a more negative drag value at the time of the secondary lift peak. At this temporal instance, we observe flow acceleration on the lower surface, as well as a massive region of recirculation on the upper surface, enhanced by increasing values of $R_v$. We attribute a further reduction in the drag levels to the strong flow recirculation observed on the upper surface.

Regarding the influence of the vortex-gust size, the findings discussed in this section align with the current understanding of its influence: larger gust sizes translate as prolonged exposure times to unsteady conditions \cite{mmuriel2020lowRe}, allowing for larger vorticity injection from the leading edge, and higher aerodynamic responses \cite{barnes2020angle}. In addition, we observe larger vorticity production levels at the leading edge for large gust sizes, leading to the formation of a distinct LEV at high values of $R_v$. The size and influence of this LEV also increase with $R_v$, and so does the magnitude of the drag fluctuations post-impingement. We attribute this to a strong recirculation pattern observed on the upper surface that provides temporary thrust. 

\subsection{Effect of vertical offset of the vortical gust}
\label{sec:initialposition}
\begin{figure}[t!]
\centering {
\vspace{0.05cm}
{\hspace*{0.cm}\includegraphics[angle=0, width= 0.95\textwidth]{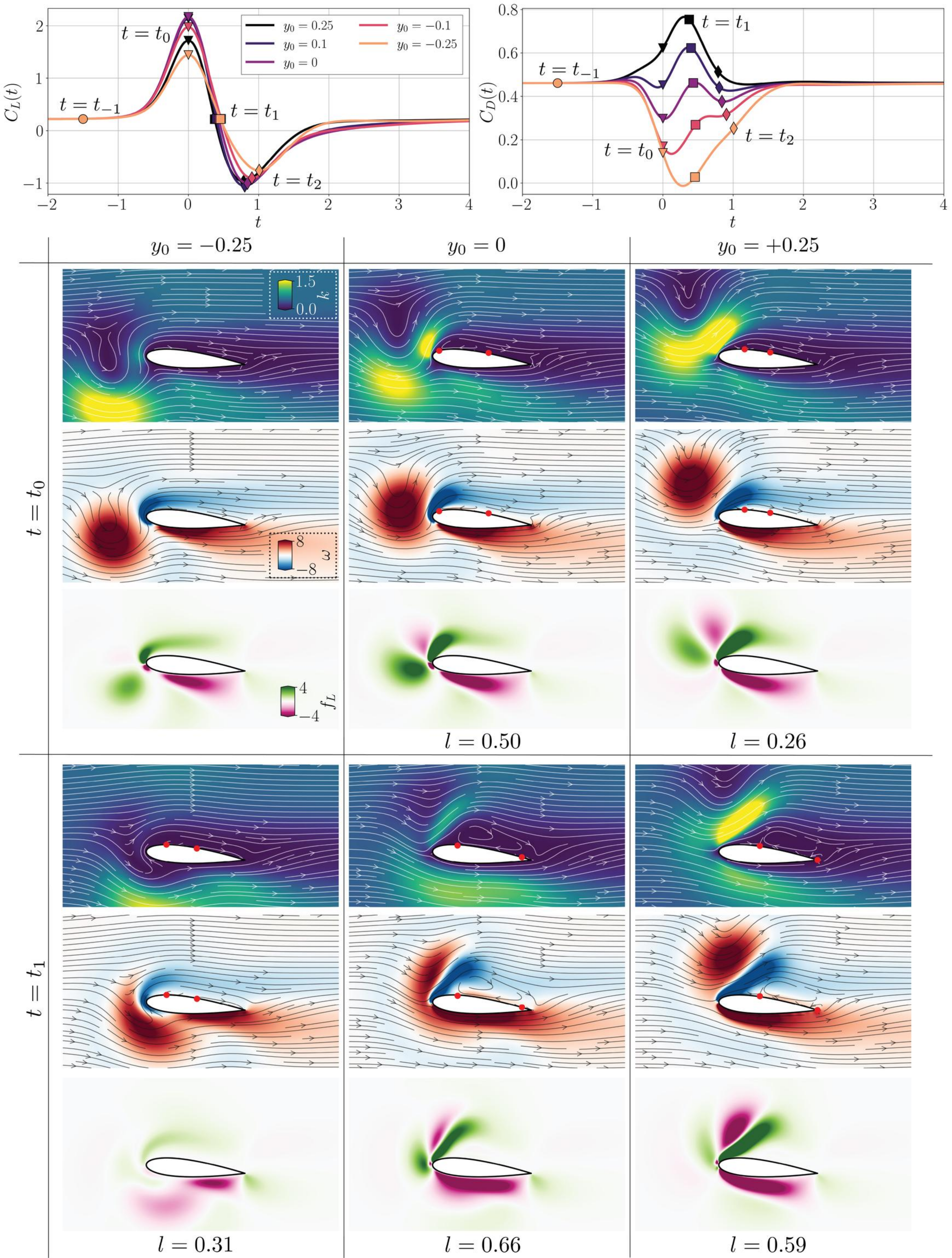}} }
\vspace{-0.cm} 
\caption{Influence of $y_0$ on $C_L(t)$ and $C_D(t)$, kinetic energy $k$, velocity (streamlines), vorticity $\omega$, and lift-element $f_L$ fields, observed during a vortex gust encounter by a NACA 0018 and $(G,R_v,\alpha)=(2,0.25,5^\circ)$.  
}
\vspace{-0.cm} 
\label{fig:results_y0}
\end{figure}

Here, we discuss the influence of the initial vertical position of the vortex gust $y_0$ with $(G,R_v,\alpha)=(2,0.25,5^\circ)$ for a NACA 0018 airfoil. The results of the analysis for various gust initial vertical positions $y_0$ are summarized in figure~\ref{fig:results_y0}. The lift curves in figure~\ref{fig:results_y0} (a) highlight a dual sensitivity to both the magnitude and sign of $y_0$: the amplitude of the first lift peak decreases with $|y_0|$ (the lift fluctuations with $|y_0|=0.1$ surpass those observed for $|y_0|=0.25$); and the peak associated with $y_0=0.1$ is greater than the peak corresponding to $y_0=-0.1$. Similarly, the lift peak observed for a given $y_0>0$ is larger than its negative counterpart.

Since $G>0$, the stagnation point is pushed toward the bottom surface, and new (negative) vorticity accumulates at the leading edge. The vertical offset of the vortex gust, however, determines the position of the stagnation point. For instance, we observe that the stagnation point moves aft along the bottom surface for $y_0=-0.25$. In contrast, in the case with $y_0=+0.25$, the stagnation point is significantly closer to the leading edge (see vorticity contours in figure~\ref{fig:results_y0}). According to the perspective of the effective angle of attack $\alpha_e$ \cite{perrotta2017transverse,biler2019transverse,menon2020gusts}, the significant recession of the stagnation point toward the leading edge for $y_0=0.25$ can be interpreted as a negative gust contribution toward the effective angle of attack, such that $\alpha_e<\alpha$. This would explain the reduced lift response observed for $y_0=0.25$ with respect to $y_0=0$ or $y_0=-0.1$. Meanwhile, while the stagnation point moves aft for $y_0=-0.25$, which would in theory signal an increase in the effective angle of attack $\alpha_e>\alpha$, the kinetic energy of the flow traveling around the leading edge is suppressed compared to the configurations shown in figures~\ref{fig:vortProd_intro} and \ref{fig:results_y0}, resulting in a reduced lift response. 

The direction of the streamlines is substantially affected by $y_0$ as well: they tightly wrap around the leading edge with no visible LSB for $y_0=-0.25$, while the streamlines lift away from the surface and an LSB of $l=0.26$ is observed for $y_0=+0.25$. Notice that an LSB was observed near the leading edge with $l=0.35$ for $y_0=-0.1$ (see figure~\ref{fig:vortProd_intro}), and that the extent of this region expands as $|y_0|$ further decreases. In particular, it expands to $l=0.50$ for $y_0=0$.

Variations of $y_0$ also yield a rich set of post-impingement dynamics. For instance, we do not observe vortex roll-up for $y_0=-0.25$. On the other hand, we observe the formation of a LEV for $y_0=0$, indicative of a stronger pressure gradient and vorticity accumulation at the leading edge. While vorticity shedding from the upper surface is observed post-impingement for $y_0=0.25$, it leads to the formation of a vortical structure with an elongated shape. Moreover, we observe a small separated region $l=0.31$ with $y_0=-0.25$ at $t=t_1$. In the other two cases, the initial LSB has shifted downstream on the upper surface and towards the trailing edge. The extent of the separated region becomes $l=0.66$ for $y_0=0$, and reaches the trailing edge with $l=0.59$ for $y_0=0.25$. In terms of lift-element analysis, we observe a faint positive lift element at the trailing edge for $y_0\leq 0$ that becomes weaker as $y_0$ increases. Moreover, the influence of $y_0$ on the magnitude of the second lift peak is less pronounced than for the first peak: although the vorticity and lift element fields exhibit different topologies, the net aerodynamic effect remains comparable. In particular, we observe a predominant negative lift element near the trailing edge for $y_0=-0.25$, whereas for $y_0=+0.25$ we observe strong positive and negative lift elements situated near the leading edge. 

The influence of $y_0$ is reflected in the drag transients as well. Notably, we observe the largest negative drag fluctuation for $y_0=-0.25$, the largest positive fluctuation for $y_0=0.25$, and the rest of the drag curves fall in between. We attribute the negative drag transients in this case to the strong local flow acceleration around the leading edge observed for $y_0=-0.25$, opposing the direction of the free-stream. As $y_0$ increases and becomes more positive, the region of local flow acceleration shifts to the upper surface and aligns with the direction of the free-stream, resulting in a heightened (positive) drag response. 

As previously reported in the literature, the magnitude of the aerodynamic responses is accentuated for small $|y_0|$ \cite{wilder1998parallel,Barnes2018Re}, and the magnitude of the lift fluctuation follows an asymmetric distribution with respect to $y_0=0$ that is dependent on the sign of $G$ \cite{horner1993asymmetric,Peng2017asymmetric}: \textit{e.g.~}we observe a larger lift fluctuation for $y_0=+0.25$ than $y_0=-0.25$. Indeed, we observe higher kinetic energy values at the leading edge for $y_0>0$, which lead to enhanced vorticity production levels in this region and a larger lift response than their negative counterpart. Lift-element analysis also reveals a stronger positive lift element at the leading edge for $y_0=+0.25$ when impingement occurs, which decreases in size and strength for $y_0=-0.25$. We additionally provide insight regarding the influence of the parameter $y_0$ on the magnitude and sign of the drag transients. 

\subsection{Effect of angle of attack}
\label{sec:angleAttack}
\begin{figure}[t!]
\centering {
\vspace{0.05cm}
{\hspace*{0.cm}\includegraphics[angle=0, width= 0.95\textwidth]{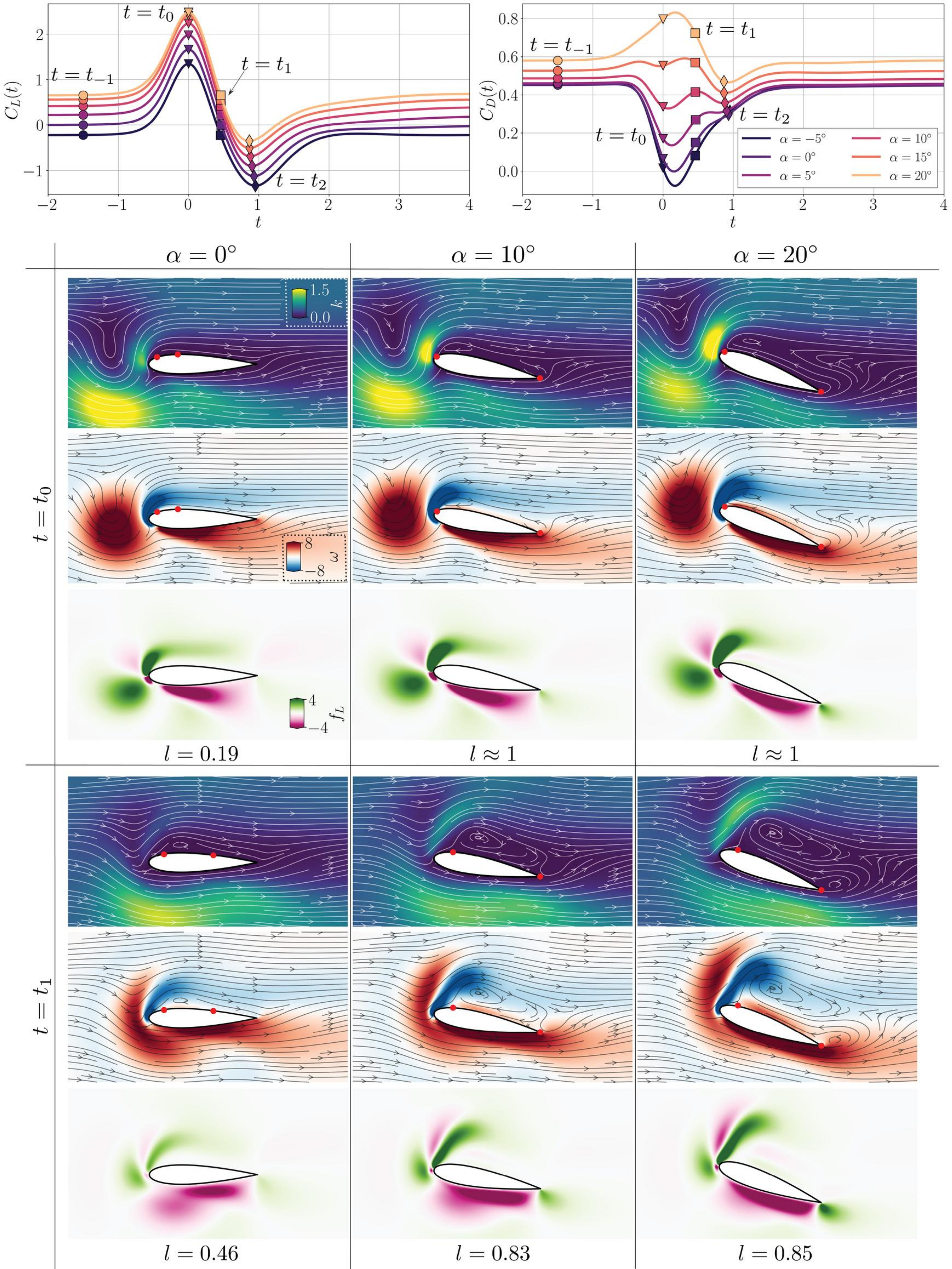}} }
\vspace{-0.cm} 
\caption{Influence of $\alpha$ on $C_L(t)$ and $C_D(t)$, kinetic energy $k$, velocity (streamlines), vorticity $\omega$, and lift-element $f_L$ fields, observed during a vortex gust encounter by a NACA 0018 and $(G,R_v,y_0)=(2,0.25,-0.1)$.  
}
\vspace{-0.cm} 
\label{fig:results_alpha}
\end{figure}

Let us now examine the influence of the angle of attack $\alpha$ on the dynamics observed during a vortex-gust encounter. Representative cases for a NACA 0018 profile with $(G,R_v,y_0)=(2,0.25,-0.1)$ are presented in figure~\ref{fig:results_alpha} at different angles of attack. At first glance, the transient component of the lift curves showcases a modest dependence on $\alpha$, although the maximum lift peak increases with this parameter. This suggests that a gust of the same properties, namely strength, size, and vertical offset, will provide the airfoil with additional circulation of similar magnitude in all cases. In other words, the unsteady effective angle of attack increases at higher incidences, but the offset with respect to $\alpha$ remains fairly consistent across configurations. We do report, however, a flattening in the magnitude of the first lift peak past $\alpha=10^\circ$, signaling the occurrence of stall.  

At $t=t_0$, we observe a small LSB with $l=0.19$ for the case $\alpha=0^\circ$. At higher incidences, $\alpha\geq 10^\circ$, we the airfoil experiences full separation ($l\approx 1$). We also report larger values of kinetic energy of the flow moving around the leading edge at high angles of attack, leading to an enhanced accumulation of vorticity at the leading edge, and a stronger positive lift element. 

As the vortex gust progresses past the leading edge, the initial LSB shifts downstream. At low angles of attack, in which no secondary separated region was observed, the LSB grows in size and reaches a length of $l=0.46$ for $\alpha=0^\circ$ at $t=t_1$. For higher angles of attack, however, the extent of the separated region reaches a value of $l=0.83$ for $\alpha=10^\circ$ and $l=0.85$ for $\alpha=20^\circ$ at $t=t_1$. Moreover, at $\alpha=20^\circ$, the flow still experiences considerable acceleration at the leading edge, as revealed by the kinetic energy distribution. As the angle of attack increases, the negative vortical structures diffused from the leading-edge post impingement become increasingly more compact, and evolve into a distinct LEV at $\alpha=20^\circ$. Moreover, we report the emergence of a positive lift element near the trailing edge that arises at high incidences, and is present both before and after impingement. 

Regarding the evolution of the drag transients, we observe the largest negative drag peak for $\alpha=-5^\circ$ and the largest positive one for $\alpha=20^\circ$. The rest of the profiles lie in between those two curves. We attribute the progressive increase in the magnitude of the drag fluctuation toward positive values to the increasing length of the gust-induced separation region observed on the upper surface at high incidences. Notice that at $\alpha=20^\circ$, the flow is fully separated, and results in high (positive) drag levels. On the other hand, we observe a small separation region for $\alpha=-5^\circ$ that remains in the form of an LSB throughout the interaction. We observe greater flow acceleration around the leading edge at high incidences as well, in alignment with the direction of the streamlines before impingement, and therefore resulting in a large (positive) increase in drag levels with respect to the unperturbed state. 

As was previously reported in the literature, unsteady stall conditions can be reached during a vortex-airfoil interaction \cite{barnes2020angle}. Moreover, with the exception of the case $\alpha=-5^\circ$, all lift curves appear to be self-similar \cite{Engin2018}, and the duration of the vortex-airfoil interaction remains fairly consistent across $\alpha$ \cite{mmuriel2020lowRe}. Although we report higher kinetic energy levels at the leading edge and high angles of attack, which would usually result in larger vorticity production levels and lift values, the lift fluctuations remain consistent at high incidences $\alpha>10^\circ$. We associate this observation with the occurrence of dynamic stall. The angle of attack has a significant influence on the drag transients as well, for which we observe positive peaks at high angles of attack and negative peaks at low angles of attack. 

\subsection{Effect of the airfoil geometry}
\label{sec:airfoil}
This subsection presents a characterization of the influence of the airfoil thickness and camber on the dynamics observed in a vortex-gust encounter. The dynamics discussed so far have been centered around a symmetric NACA 0018 profile. We first discuss the effect of varying the airfoil thickness $\tau$ on the trends discussed so far for symmetrical airfoils in \S\ref{sec:thickness}. The influence of camber $\eta$ is then discussed in \S\ref{sec:camber} for thin airfoils ($0.06$ thickness) to highlight the influence of the camber. 

\subsubsection{Effect of airfoil thickness}
\label{sec:thickness}
\begin{figure}[t!]
\centering {
\vspace{0.cm}
{\hspace*{-0.cm}\includegraphics[angle=0, width= 0.95\textwidth]{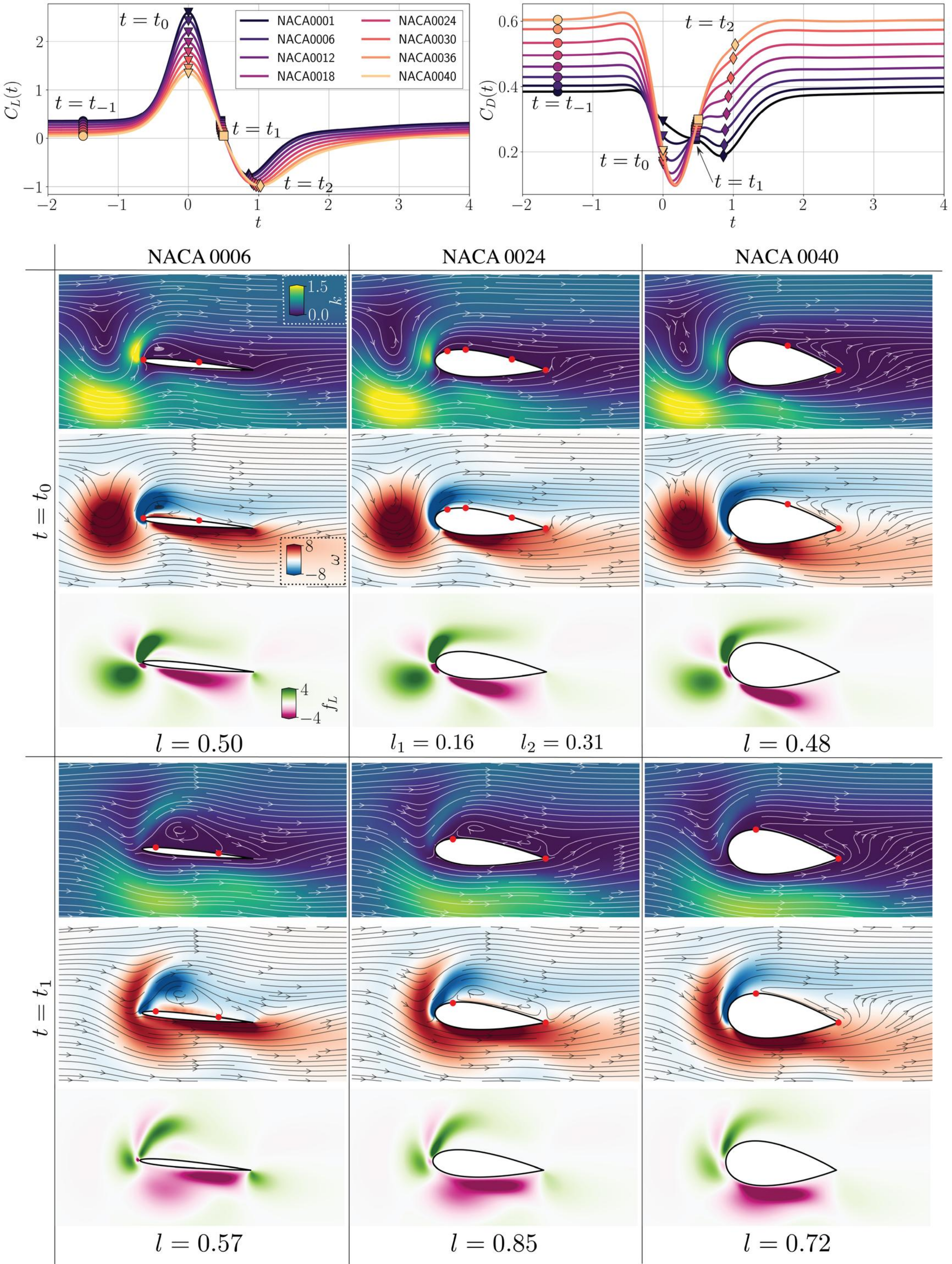}} }
\vspace{-0.cm} 
\caption{Influence of $\tau$ on $C_L(t)$ and $C_D(t)$, kinetic energy $k$, velocity (streamlines), vorticity $\omega$, and lift-element $f_L$ fields, observed during a vortex gust encounter with $(G,R_v,y_0,\alpha)=(2,0.25,-0.1,5^\circ)$.  
}
\vspace{-0.cm} 
\label{fig:results_thickness}
\end{figure}

We consider symmetric airfoils of varying thickness $\tau$
at $\alpha=5^\circ$ with $G=2$, $R_v=0.25$ and $y_0=-0.1$, as shown in figure~\ref{fig:results_thickness}. Two main trends are drawn from these results: an increasing airfoil thickness attenuates the lift response, and we observe negative drag peaks for all airfoils of increasing magnitude for increasingly thick profiles.

Let us first discuss the vortex dynamics preceding vortex impingement. First, we observe that the kinetic energy decreases at the leading edge for thick airfoils. An LSB of extent $l=0.50$ emerges near the leading edge for $\tau=0.06$, and is significantly reduced to $l=0.16$ for $\tau=0.24$ (NACA 0024). Nonetheless, an additional separated region arises in this case from the trailing edge on the upper surface and has an extent of $l=0.31$. As airfoil thickness further increases, we observe a separated region arising from the trailing edge with a length of $l=0.48$. 

The high accumulation of vorticity at the leading edge for thin profiles leads to a higher diffusion flux of vorticity to enter this flow, facilitating the development of a LEV once the pressure gradient subsides, as observed for $\tau=0.06$ (NACA 0006). This is not observed for thicker profiles, as there is not enough vorticity accumulation at the leading edge to promote the development of a LEV. As the lower and upper portions of the gust advance downstream, the LSB observed for $\tau=0.06$ expands to a length of $l=0.57$. Conversely, for thick profiles, the extent of the secondary separated region increases as well post-impingement, reaching a length of $l=0.72$ for $\tau=0.40$ (NACA 0040). Notably, for $\tau=0.24$, the LSB appears to overlap with the secondary recirculation region at this time, and the total extent of the separated region amounts to $l=0.85$.

\begin{figure}[t!]
\centering {
\vspace{0.cm}
{\hspace*{0.cm}\includegraphics[angle=0, width= 0.8\textwidth]{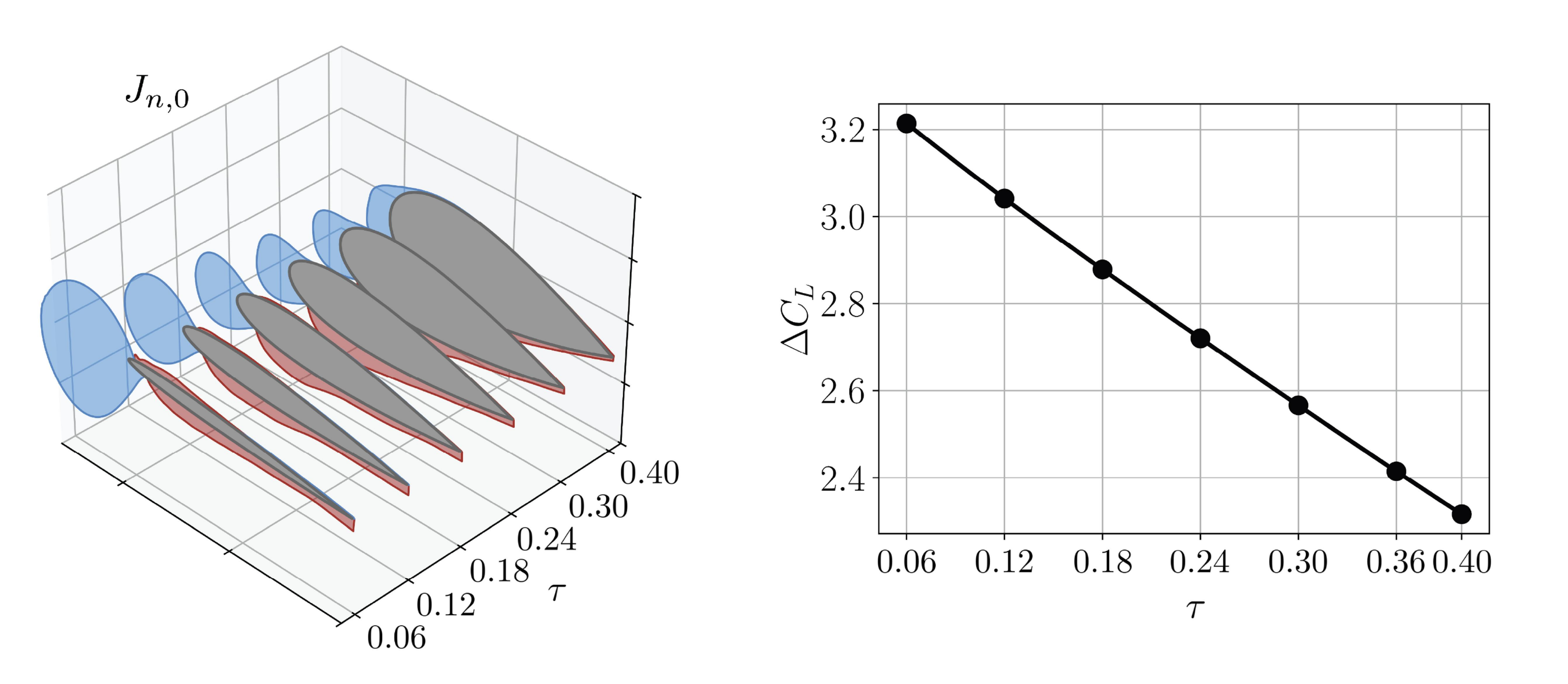}} }
\vspace{-0.cm} 
\caption{(Left) Vorticity production flux $J_{n,0}(t_0)$ at $t=t_0$ (impingement) in which red and blue shades denote regions of positive and negative values; and (right) lift fluctuation $\Delta C_L$, against airfoil thickness during a gust-airfoil interaction with $(G,R_v,y_0,\alpha)=(2,0.25,-0.1,5^\circ)$. 
}
\vspace{-0.cm}
\label{fig:diagram_curvature}
\end{figure}

The vorticity production levels observed at impingement for symmetric airfoils are shown in figure~\ref{fig:diagram_curvature}(left). We report significantly higher vorticity production levels from the leading edge for thinner airfoils, in agreement with the heightened lift response recorded for such profiles. A linear trend is observed between the lift fluctuation $\Delta C_L$ and $\tau$ in figure~\ref{fig:diagram_curvature}(right), further underscoring the relevance of the curvature term $\kappa$ on the total vorticity production (see second term in \ref{eq:smomentum}). This drastic reduction in $\kappa$ observed for thicker airfoils, following $R_c=1.1019 \tau^2$ in \cite{abbott1959theory} for four-digit NACA airfoils, translates into a mitigated influence of the curvature term in the vorticity production flux, leading to a more modest lift response.

The influence of the airfoil geometry is also reflected in the drag transients. While the lift fluctuation is attenuated for thick airfoils, the magnitude of the drag fluctuation is enhanced by thick profiles. We attribute this difference to the enlarged perimeter of thick airfoils: the gust-induced increment in the effective angle of attack is agnostic to the airfoil shape, and the region of flow acceleration under the leading edge, which opposes the direction of the free stream, produces an enhanced drag reduction for airfoils with a larger perimeter until $t_0$. Interestingly, we observe larger (more negative) drag values for thin airfoils at $t_2$, which we attribute to the thrust provided by the strong separation bubble observed on the upper surface and induced by the LEV post-detachment as it convects downstream. Note that for thick airfoils, we merely observe a separated region of great extent instead. 

Previous studies had hinted at the possibility of mitigating the lift response in gusty environments using blunt leading edges \cite{Gementzopoulos2024blunt}, or thick airfoils in general \cite{zhong2024geometrytransonic}. Here, we have examined the influence of airfoil geometry from the perspective of vorticity production fluxes and determined that the gust-induced vorticity fluxes at the leading edge are effectively attenuated for low-curvature airfoils. Furthermore, we report a linear decrease in the magnitude of the lift fluctuation with $\tau$. The influence of the airfoil thickness is also reflected in the drag transients: thick airfoils exhibit larger (more negative) drag levels before impingement, while thin airfoils experience larger drag levels after impingement due to the influence of the vertical structure that rolls up on the upper surface post-impingement.

\subsubsection{Effect of airfoil camber}
\label{sec:camber}
\begin{figure}[t!]
\centering {
\vspace{0.cm}
{\hspace*{0.cm}\includegraphics[angle=-0, width= 0.95\textwidth]{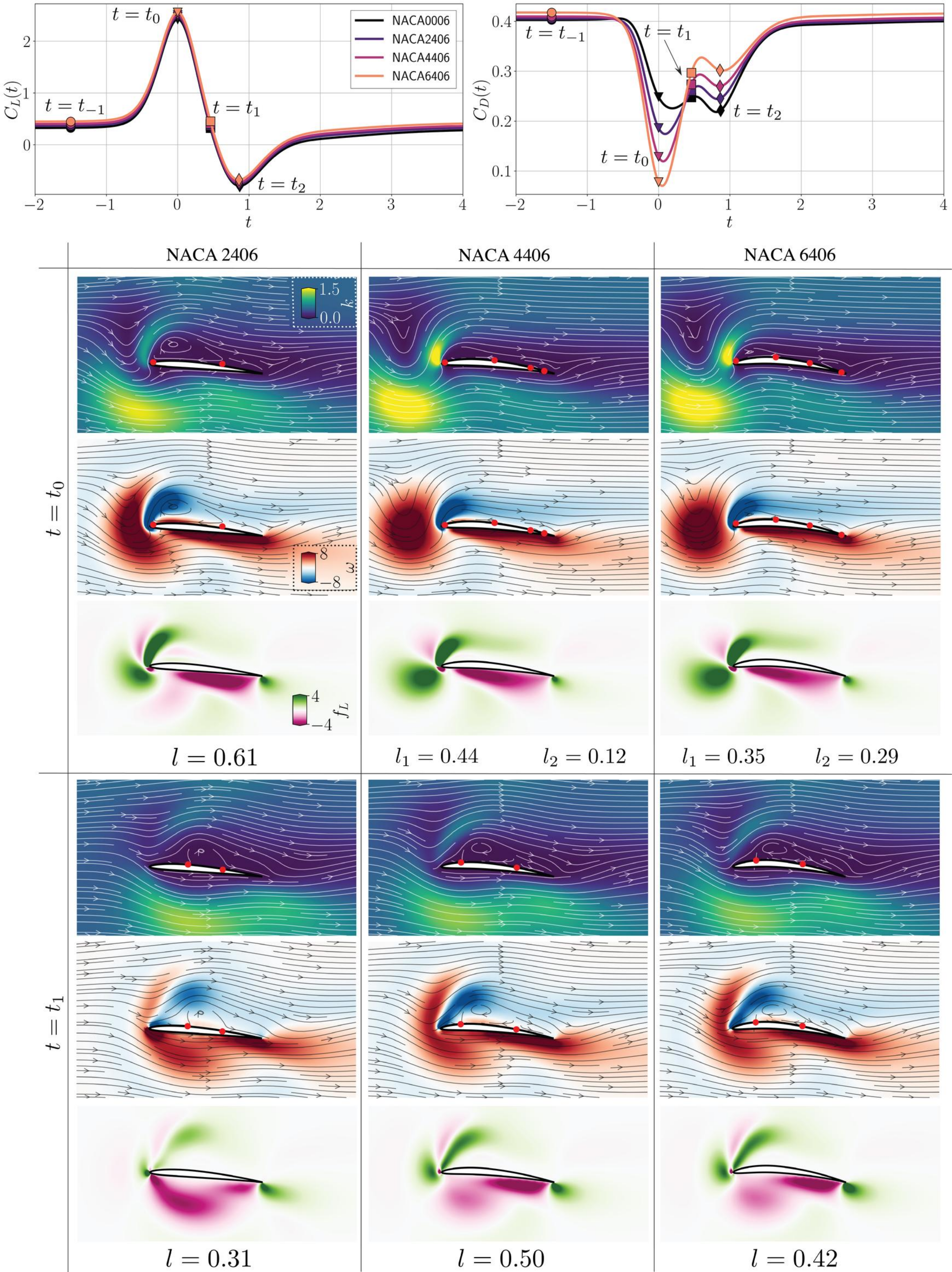}} }
\vspace{-0.cm}
\caption{Influence of $\eta$ on $C_L(t)$ and $C_D(t)$, kinetic energy $k$, velocity (streamlines), vorticity $\omega$, and lift-element $f_L$ fields, observed during a vortex gust encounter with $(G,R_v,y_0,\alpha;\xi)=(2,0.25,-0.1,5^\circ;0.4)$.  
}
\vspace{-0.cm} 
\label{fig:results_camber}
\end{figure}

Next, let us characterize the influence of the airfoil camber, which has not been considered thus far. Here, we examine the differences in the aerodynamic response of the following profiles: $\{$NACA 0006, NACA 2406, NACA 4406, NACA 6406$\}$. All airfoils share the same airfoil thickness $\tau=0.06$, and the location of maximum camber is fixed at $\xi=0.4$. This dataset is thus parametrized according to the camber magnitude $\eta$, with values $\eta=\{0,0.02,0.04,0.06\}$. The corresponding results are presented in figure~\ref{fig:results_camber}. 

Camber effects seem to have little influence on the magnitude of the lift fluctuations in this case, while the influence on the drag is significantly more noticeable. The flow states observed at impingement reveal increased levels of kinetic energy at the leading edge for increasing camber. On the one hand, we observe an LSB for all three profiles at $t=t_0$. In particular, the extent of this region increases from $l=0.50$ to $0.61$ between $\eta=0$ and $\eta=0.02$ ($\eta=0$ shown in figure~\ref{fig:results_thickness}). Moreover, while the extent of the LSB reduces for $\eta>0.02$, it is compensated by a growing separated region that originates from the trailing edge with a positive lift contribution (as revealed by force-element analysis). 

After impingement, at $t=t_1$, the separated region that originated from the trailing edge collapses (although the positive lift element prevails at the trailing edge), and the LSB moves downstream along the upper surface. Interestingly, the extent of this separated region decreases with respect to $\eta=0$ for all cambered airfoils. We report the formation of a LEV post-impingement that becomes weaker as the magnitude of the camber decreases, signaling a strong accumulation of vorticity at the leading edge pre-impingement for airfoils with high camber. We also observe boundary-layer thickening on the pressure side that becomes more concentrated at the trailing edge with increasing $\eta$.

\section{Characterizing the effect of multiple parameters on the nonlinear dynamics observed in gust encounters}
\label{sec:results_multi}
In this manuscript, the discussions have focused on the effect of an isolated airfoil or gust parameter. In fact, coupling effects between two or more variables often manifest during gust-airfoil interactions. Here, we include a discussion of the influence of the variables in our parameter space to consider the following combined effects, although the individual influence of each parameter remains consistent across different parameter configurations. The combined influence of angle of attack ($\alpha$) and airfoil thickness ($\tau$) is discussed in \S\ref{sec:angleThickness}, and the influence of angle of attack, gust ratio ($G$), and initial vertical position of the vortex gust ($y_0$) is presented in in \S\ref{sec:angleGustInitial}. 

\subsection{Combined effect of angle of attack and airfoil thickness}
\label{sec:angleThickness}
\begin{figure}[t!]
\centering {
\vspace{-0.cm}
{\hspace*{0.cm}\includegraphics[angle=0, width= 1\textwidth]{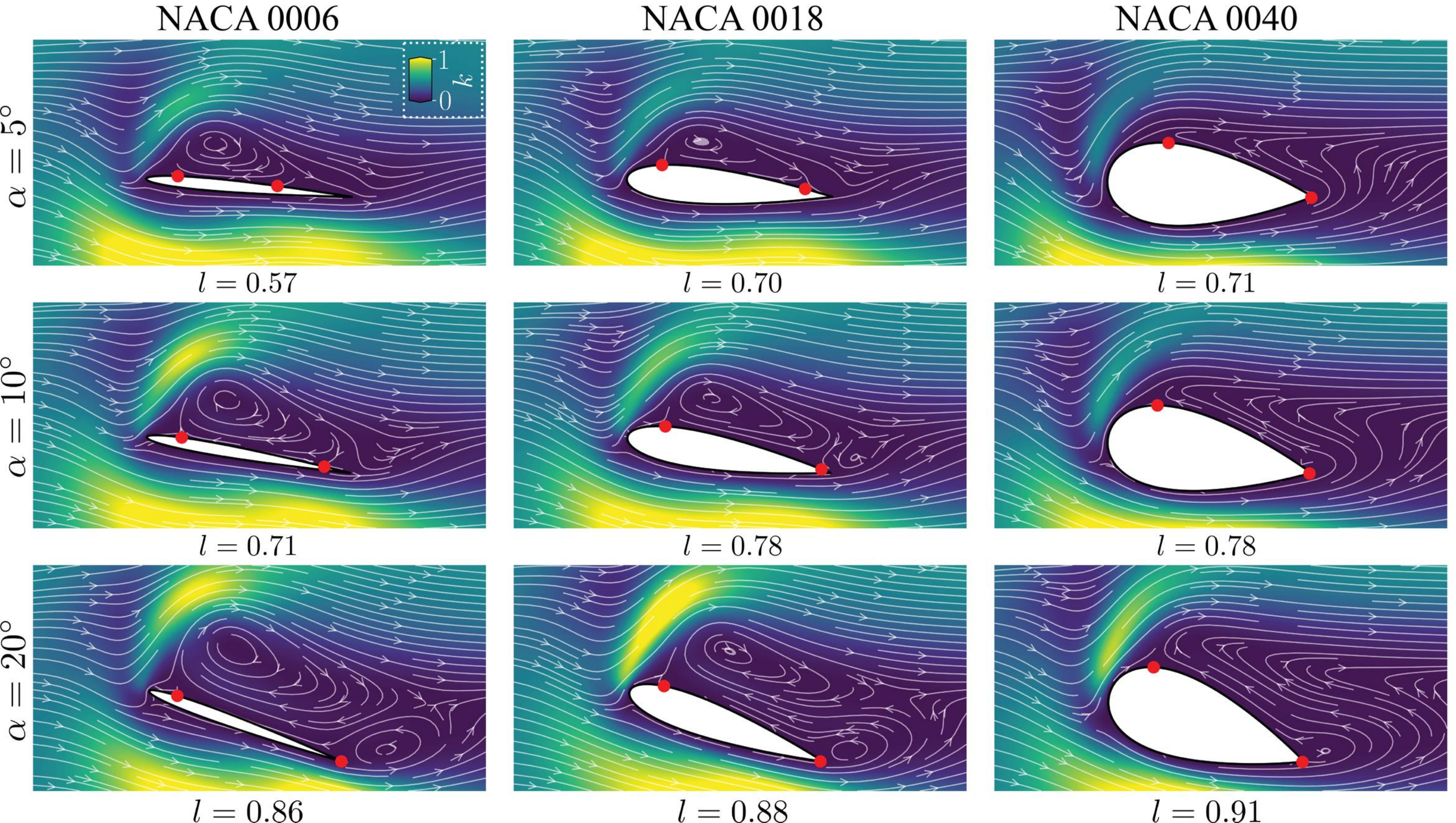}} }
\vspace{-0.4cm}
\caption{Kinetic energy ($k$) and velocity (streamlines) fields at $t=t_1$ for a set of symmetrical airfoils of various thicknesses $\tau$ at different angles of attack $\alpha$ with $(G,R_v)=(2,0.25)$.
}
\vspace{-0.3cm}
\label{fig:thicknessvsAlphavsG}
\end{figure}

Throughout this manuscript, it has been indicated that several factors can trigger the development of a wide separated region that originates from the leading-edge and grows upstream along the upper surface. Prominent examples of this behavior include vortex gusts of large size ($R_v$), high angles of attack ($\alpha$), and thick airfoils. The combined influence of thickness and angle of attack on this separated region, previously characterized under steady conditions in \S\ref{sec:baselines}, is illustrated in figure~\ref{fig:thicknessvsAlphavsG}. For the thinnest profile, that is $\tau=0.06$, we observe a separated region at $\alpha=5^\circ$ corresponding to an LSB that originated near the leading edge before impingement. At higher angles of attack, the extent of the LSB increases, and at $\alpha=20^\circ$ we observe the overlap of a secondary separated region near the trailing edge, and the LSB, indicating stall-like conditions. Meanwhile, for the thickest profile, that is a $\tau=0.40$, we observe a separated region emerging from the trailing edge at $\alpha=5^\circ$ whose extent also increases at higher incidences. 

\begin{figure}[t!]
\centering {
\vspace{0.2cm}
{\hspace*{0.cm}\includegraphics[angle=0, width= 1\textwidth]{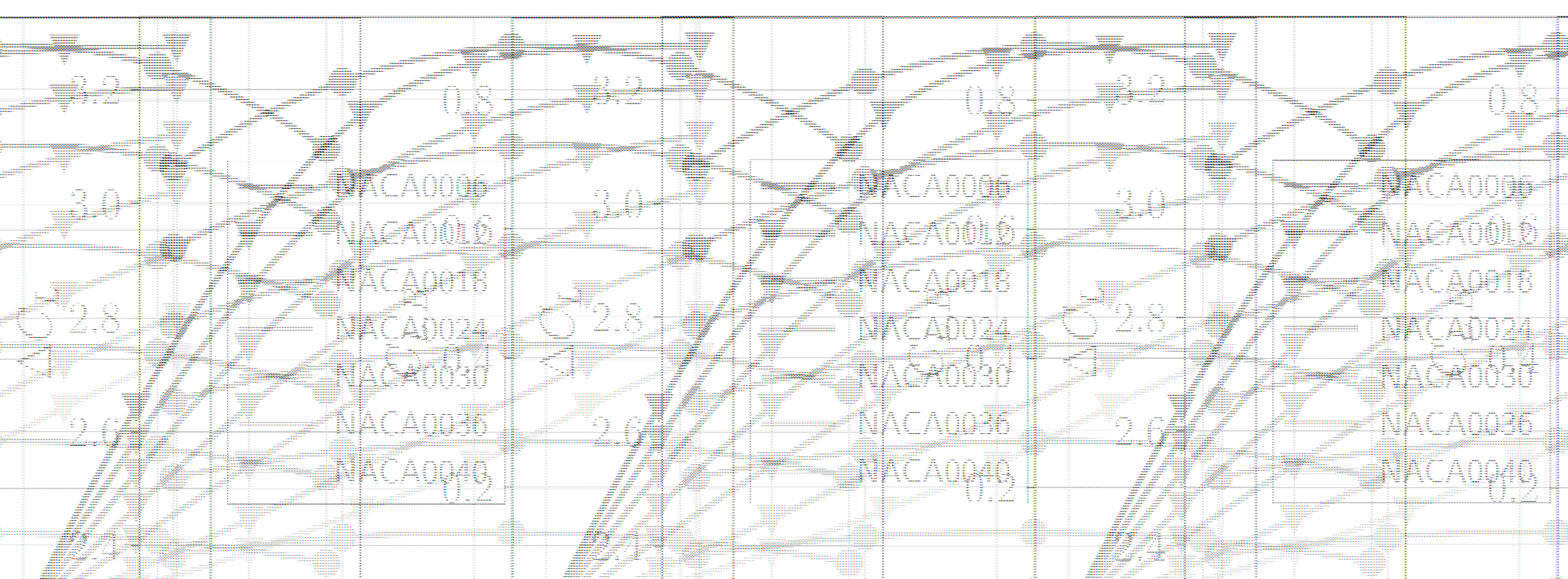}} }
\vspace{-0.6cm}
\caption{Magnitude of the lift fluctuation $|\Delta C_L|$ against $\alpha$ and $\tau$, with $(R_v,y_0)=(0.25,-0.1)$. 
}
\vspace{-0.cm} 
\label{fig:thicknessvsAlphavsG_trends}
\end{figure}

In general, thick airfoils experience reduced (or mild) lift fluctuations during a gust encounter due to the suppression of the vorticity production levels at the leading edge (see figure~\ref{fig:diagram_curvature}). Thin airfoils, on the other hand, experience a larger lift fluctuation because the high curvature at the leading edge promotes high vorticity production levels. Moreover, a linear relationship was found between the magnitude of the lift fluctuation and airfoil thickness in \S\ref{sec:thickness} for $(G,\alpha)=(2,5^\circ)$. Nonetheless, this relationship persists at different angles of attack and gust ratios, as indicated in figure~\ref{fig:thicknessvsAlphavsG_trends} (right). On the other hand, we observe a decrease in the magnitude of the lift fluctuations after $\alpha=10^\circ$ for all airfoils, signaling the occurrence of dynamic stall at higher incidences. 

\subsection{Combined effect of angle of attack, gust initial position, and gust ratio}
\label{sec:angleGustInitial}
This section examines the influence of the angle of attack on the trends observed in \S\ref{sec:initialposition} for various $y_0$. At low incidences, we observed the development of an LSB for $y_0>-0.25$ whose extent increased as the core of the vortex gust is initialized closer to the leading edge (\textit{i.e.}~small $|y_0|$). As $y_0$ becomes increasingly positive, the location of this separated region is found further aft on the upper surface. We identified an instance in which the LSB extended up to the trailing edge post impingement for $y_0=+0.25$. 

\begin{figure}[t!]
\centering {
\vspace{-0.cm}
{\hspace*{0.cm}\includegraphics[angle=0, width= 1\textwidth]{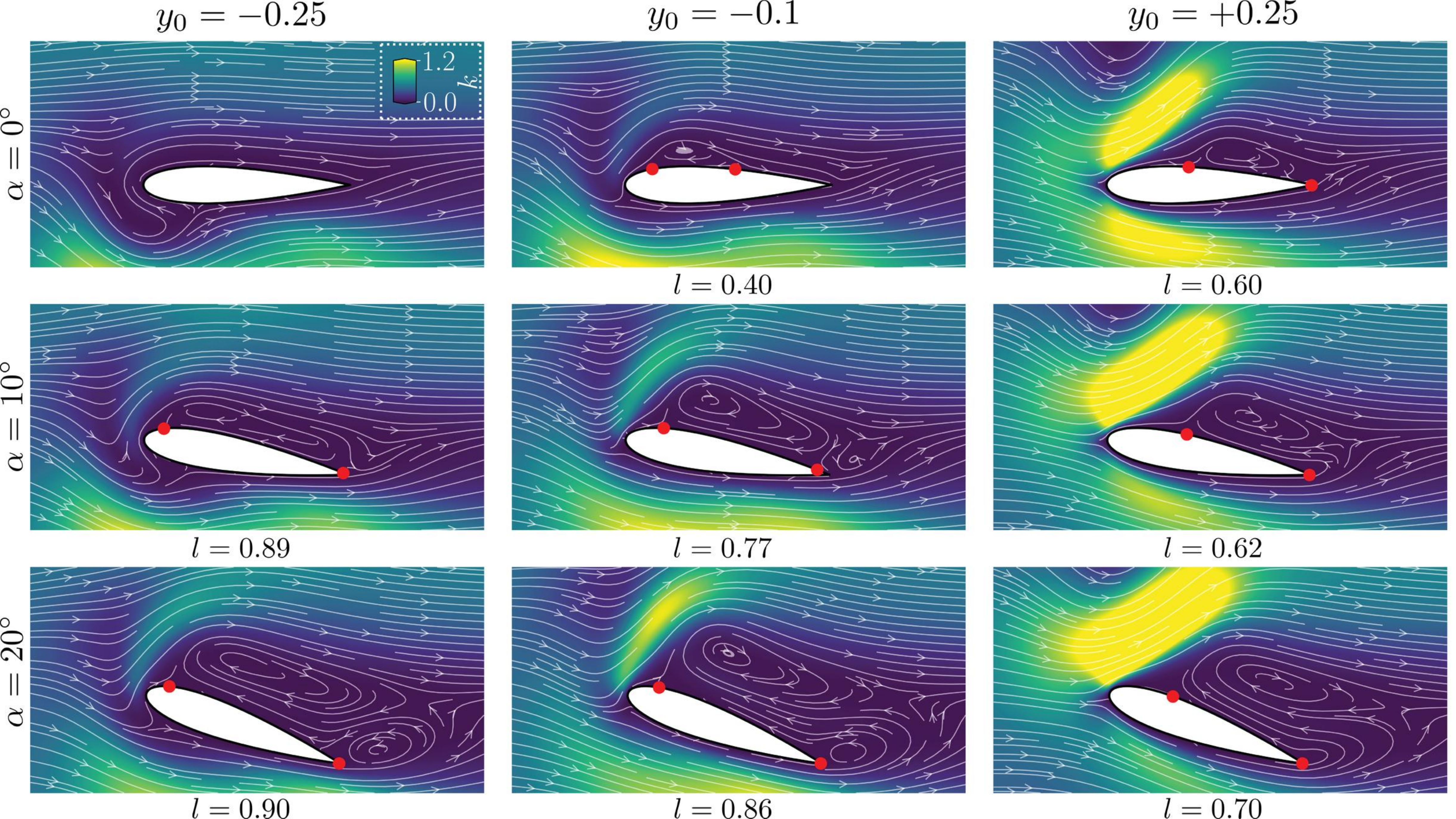}} }
\vspace{-0.3cm}
\caption{Kinetic energy ($k$) and velocity (streamlines) fields at $t=t_1$ for a NACA 0018 airfoil at different angles of attack $\alpha$ and vertical offsets $y_0$ with $(G,R_v)=(2,0.25)$.
}
\vspace{-0.cm}
\label{fig:Alphavsy0}
\end{figure}

The influence of both $\alpha$ and $y_0$ on the extent and location of the separated region at $t=t_1$ is presented in figure~\ref{fig:Alphavsy0}. For all $y_0$, we observe a consistent increase in the size of the separated region in configurations with increasing angles of attack. This increase, however, is significantly less dramatic for $y_0=0.25$. Notably, the sharpest increase is observed for $y_0=-0.25$, for which the recirculation region grows from $l=0$ at $\alpha=0^\circ$ to a full separation $l\approx 1$ at $\alpha=20^\circ$. Moreover, at $\alpha=20^\circ$, we observe a large separated region for all $y_0$ that originates at the trailing edge and expands upstream. The extent of this region, however, is the largest for $y_0=-0.25$, and progressively decreases as $y_0$ becomes more positive.  

\begin{figure}[t!]
\centering {
\vspace{0.cm}
{\hspace*{0.cm}\includegraphics[angle=0, width= 0.9\textwidth]{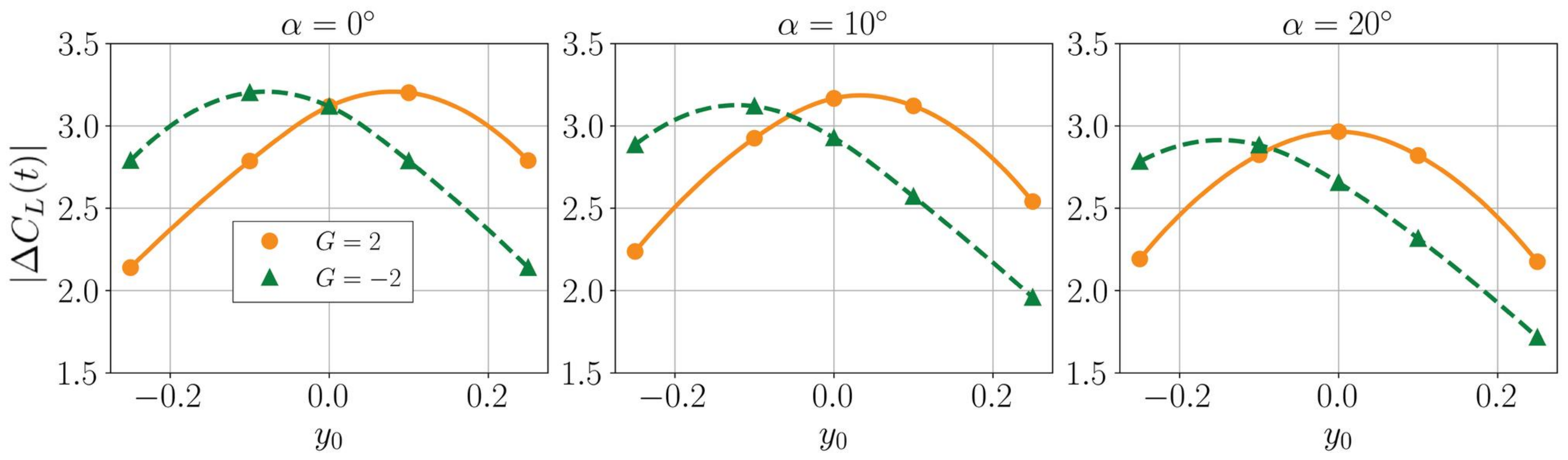}} }
\vspace{0.cm}
\caption{
Magnitude of the lift fluctuation against $y_0$ for a NACA 0018 at different angles of attack for $G=\{-2,2 \}$ with $R_v=0.25$.
}
\vspace{-0.cm}
\label{fig:schematicInitialPos}
\end{figure}

The magnitude of the lift fluctuation $|\Delta C_L(t)|$ at different angles of attack against $y_0$ is shown in the bottom row of figure~\ref{fig:schematicInitialPos}. Notably, the curves become increasingly symmetric with respect to $y_0=0$ in configurations at higher angles of attack and positive gust ratios. Meanwhile, we observe the opposite trend for $G<0$: the curve becomes more asymmetric at high incidences. In fact, this asymmetric behavior has been previously reported in \cite{horner1993asymmetric,Peng2017asymmetric} for $\alpha=0^\circ$. Furthermore, these trends can be mirrored to negative angles of attack, although not shown here, such that the curve for $G=-2$ becomes symmetric about $y_0=0$ for $\alpha=-20^\circ$, and the curve for $G=2$ becomes a mirror against $y_0=0$ of the curve for $G=-2$ at $\alpha=20^\circ$. Notably, these trends hold for the present choice of $|G|$, and would change for other gust ratios and/or parameter combinations.  

\section{Concluding remarks}
\label{sec:conclusions}
This work examined in detail the effects of airfoil and gust parameters on the nonlinear dynamics observed during extreme gust-airfoil interactions. A collection of vortex-gust encounters was compiled at a chord-based Reynolds number $Re_c=100$ for stationary airfoils. The parameter space of interest includes the gust ratio $G$, gust radius $R_v$, gust initial vertical position $y_0$, angle of attack $\alpha$, airfoil thickness $\tau$, and airfoil maximum camber $\eta$. While the influence of several parameters has been established in previous studies, the contribution of airfoil geometry to gust-airfoil interactions has not been fully explored. 

The individual influence of each parameter was discerned from a systematic examination of different gust encounters, though interactions among different parameters can introduce additional nuance. In general, larger aerodynamic responses are associated with larger (high $R_v$) and stronger gusts (high $|G|$), driven by an accentuated vorticity flux from the leading edge. The magnitude of the response also depends on the angle of attack, the initial vertical position of the vortex gust, and the sign of the vortex gust. In practical terms, lift fluctuations can be attenuated by increasing the distance between the airfoil and the gust. Moreover, at low incidences, the response is attenuated when passing below positive (counterclockwise) gusts and above negative (clockwise) gusts. The dependence on the sign of $y_0$ subsides at high incidences, for which collisions with $y_0=0$ are the most magnified. Overall, the observations regarding these parameters align well with and complement previously documented studies, while adding perspective from the standpoint of vorticity production and the interpretation of lift-element fields, and offering new insight into the behavior of drag transients. 

The greatest contribution of this work lies in elucidating the effect of airfoil geometry. In particular, airfoil thickness plays a critical role. A systematic examination of airfoils of increasing thickness reveals monotonic trends in the vorticity flux from the leading edge, suggesting a potential avenue for lift attenuation in the context of airfoil design. More specifically, thick airfoils with blunt leading edges (low curvature) reduce the pressure gradients and aerodynamic responses during a gust encounter, whereas thin airfoils with sharp leading edges have the opposite effect. 

The characterizations and insights provided in this work constitute a significant contribution to the broader understanding of the complex, nonlinear vortex dynamics that govern gust-airfoil interactions, laying the groundwork for the development of lift attenuation techniques by leveraging the knowledge of gust response characteristics and the influence of airfoil thickness. While thick airfoils may not offer optimal aerodynamic performance under steady conditions, strategies that mimic the behavior of such airfoils (in particular at the leading edge) in unsteady regimes could prove valuable in future investigations. 

\section*{Acknowledgements}
BLD and KT were supported by the US Department of Defense Vannevar Bush Faculty Fellowship under grant number N00014-22-1-2798. We gratefully acknowledge the technical support and insightful discussions with Kai Fukami, Hiroto Odaka, Vedasri Godavarthi, and Victoria Rolandi. 

\section*{Declaration of interest}
The authors report no conflict of interest.

\appendix
\section{Mesh convergence studies}
\label{app:convergence}
As indicated in \S\ref{sec:ibpm}, the (regular) mesh used throughout this work, consists of 5 subdomains, the finest one spanning between $-4 \leq x \leq 5$ and $-2 \leq y \leq 3$ with $M=1152$ and $N=640$ cells, respectively. The total extent of the computational domain is $L_x=72$ in $x$ and $L_y=40$ in $y$. 

\begin{figure}[t!]
\centering {
\vspace{0.cm}
{\hspace*{0.cm}\includegraphics[angle=0, width= 0.95\textwidth]{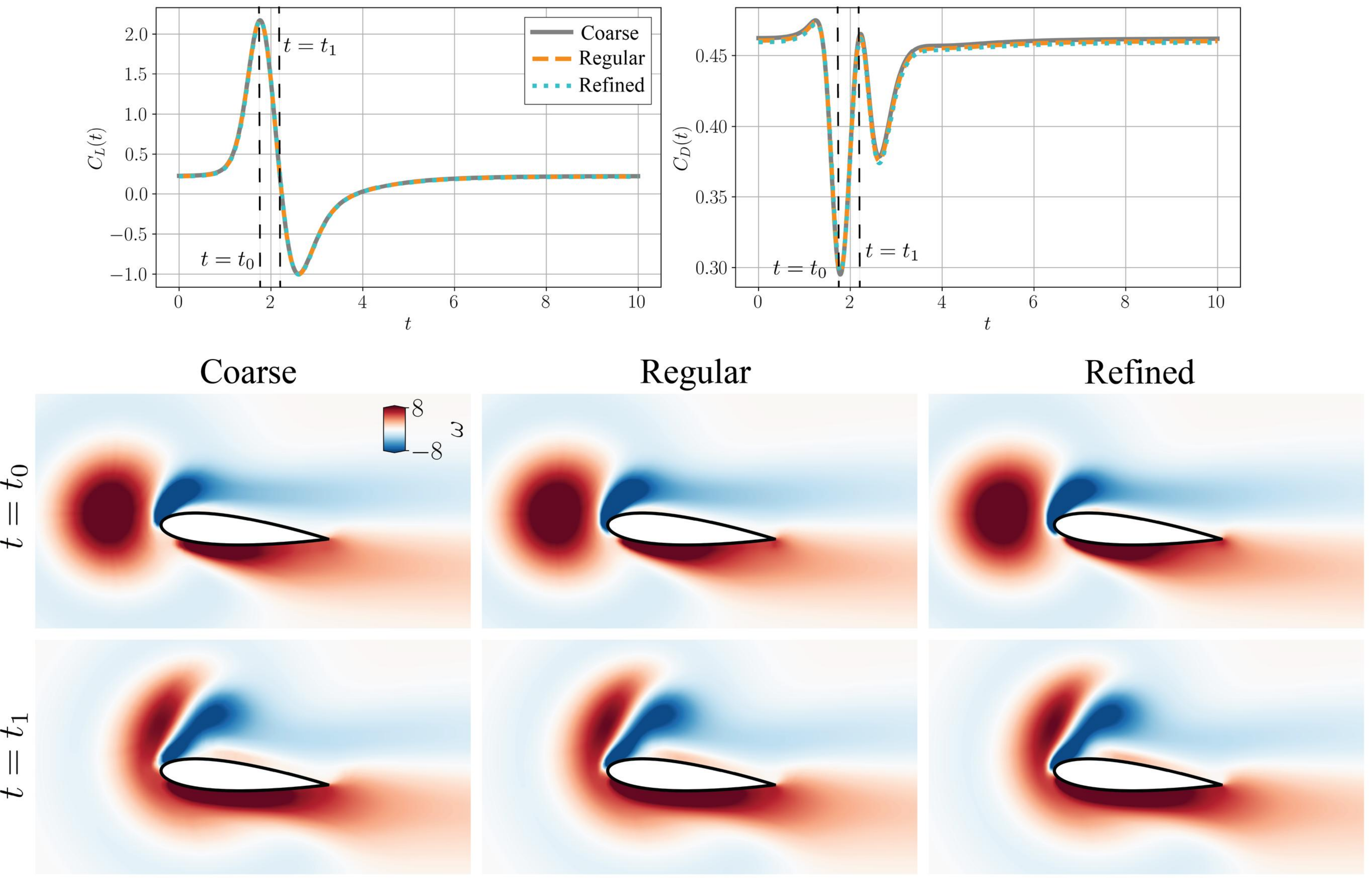}} }
\vspace{-0.3cm}
\caption{Comparison between $C_L(t)$ and $C_D(t)$, and vorticity $\omega$ fields with three different meshes observed during a vortex gust encounter by a NACA 0018 and $(G,R_v,y_0,\alpha)=(2,0.25,0,5^\circ)$.  
}
\vspace{-0.cm} 
\label{fig:convergence}
\end{figure}

To asses mesh convergence, two additional meshes were considered within the same spatial domain. The first, referred to as ``coarse", has $M=768$ and $N=424$ cells, while the second, referred to as ``refined", has $M=1536$ and $N=860$ cells. The aerodynamic coefficients, as well as the vorticity fields at $t=t_0$ and $t=t_1$ observed during a vortex gust encounter by a NACA 0018 and $(G,R_v,y_0,\alpha)=(2,0.25,0,5^\circ)$ are shown in figure~\ref{fig:convergence}. Across all three meshes, the aerodynamic coefficients and flow fields show close agreement, supporting the validity of the regular grid for this study.

\begin{figure}[t!]
\centering {
\vspace{0.cm}
{\hspace*{0.cm}\includegraphics[angle=0, width= 0.8\textwidth]{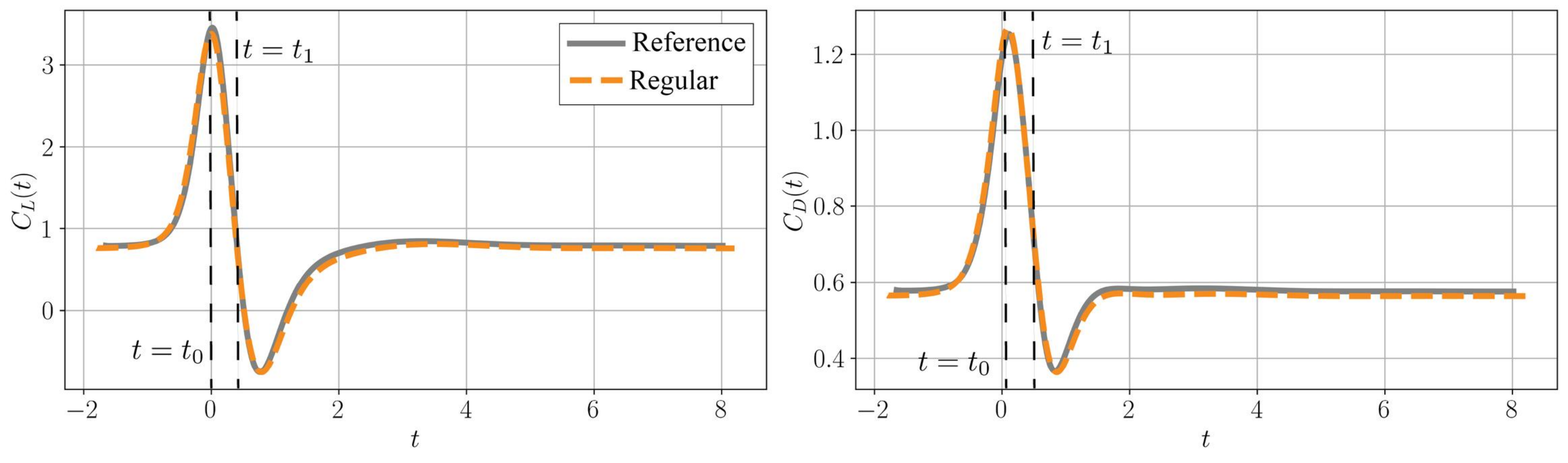}} }
\vspace{-0.3cm}
\caption{Comparison between $C_L(t)$ and $C_D(t)$ obtained with the regular mesh, against the results presented in \cite{Fukami2024lift}, observed during a vortex gust encounter by a NACA 0012 and $(G,R_v,y_0,\alpha)=(2.6,0.25,0,20^\circ)$.  
}
\vspace{-0.cm} 
\label{fig:convergence_kai}
\end{figure}

Further validation is provided in figure~\ref{fig:convergence_kai}, which presents the lift and drag coefficients against the results described in \cite{Fukami2024lift}, for a NACA 0012 with $(G,R_v,y_0,\alpha)=(2.6,0.25,0,20^\circ)$. In addition, table~\ref{tab:convergence_kai} summarizes the baseline lift and drag coefficients, as well as the lift and drag fluctuations, for both the present regular mesh and those obtained by \cite{Fukami2024lift}. We observe a good agreement between the coefficients, further confirming the adequacy of the present regular mesh. 

\begin{table}
  \begin{center}
\def~{\hphantom{0}}
  \begin{tabular}{lcccc}
      Mesh & $C_{L,b}$  & $C_{D,b}$  &  $\Delta C_L$ & $\Delta C_D$ \\
      \hline
      Fukami et al. \cite{Fukami2024lift} (reference) & 0.789  & 0.578 & 4.207 & 0.887 \\
      Regular & 0.761 & 0.565 & 4.135 & 0.901 \\
  \end{tabular}
  \caption{Validation of regular mesh by comparison of the present baseline lift and drag coefficients, along with lift and drag fluctuations in a vortex gust encounter by a NACA 0012 and $(G,R_v,y_0,\alpha)=(2.6,0.25,0,20^\circ)$ described in \cite{Fukami2024lift}. }
  \label{tab:convergence_kai}
  \end{center}
\end{table}


\bibliographystyle{unsrt}
\bibliography{Master}

\end{document}